%% file: template.tex
\documentclass{article}

\usepackage{arxiv}

\usepackage[T1]{fontenc}    
\usepackage{hyperref}       
\usepackage{url}            
\usepackage{booktabs}       
\usepackage{amsfonts}       
\usepackage{nicefrac}       
\usepackage{microtype}      
\usepackage{lipsum}		
\usepackage{graphicx}
\usepackage{natbib}
\usepackage{doi}

\usepackage{makecell}
\usepackage[utf8x]{inputenc}
\usepackage{tfrupee}
\usepackage{adjustbox}
\usepackage{ragged2e}
\usepackage{array}
\usepackage{hyperref}
\usepackage{algorithm}        
\usepackage{algpseudocode}    
\usepackage{amsmath}          
\usepackage{microtype}        
\usepackage{etoolbox}         

\title{\textit{JEEVHITAA} - An HCAI Ecosystem to Support Collective Care}
\renewcommand{\shorttitle}{JEEVHITAA}

\author{ \href{https://orcid.org/0000-0001-5787-4858}{\includegraphics[scale=0.06]{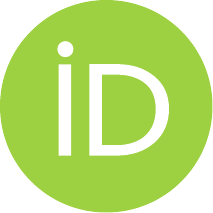}\hspace{1mm}Shyama Sastha~Krishnamoorthy Srinivasan}\\
	IIIT-Delhi\\
	New Delhi, India\\
	\texttt{shyamas@iiitd.ac.in} \\
    \And
    \href{https://orcid.org/0009-0006-6650-924X}{\includegraphics[scale=0.06]{orcid.pdf}\hspace{1mm}Harsh~Pala}\\
	IIIT-Delhi\\
	New Delhi, India\\
	\texttt{harsh24119@iiitd.ac.in} \\
    \And
    \href{https://orcid.org/0000-0002-0286-6997}{\includegraphics[scale=0.06]{orcid.pdf}\hspace{1mm}Mohan~Kumar}\thanks{Both authors are advisors of this work.}\\
	RIT, NY\\
	Rochester, New York, US\\
	\texttt{mjkvcs@rit.edu} \\
	\And
	\href{https://orcid.org/0000-0003-2152-1027}{\includegraphics[scale=0.06]{orcid.pdf}\hspace{1mm}Pushpendra~Singh}\footnotemark[1]\\
	IIIT-Delhi\\
	New Delhi, India\\
	\texttt{psingh@iiitd.ac.in} \\
}

\renewcommand{\shorttitle}{Designing for \emph{Proactive Collective Care}}

\hypersetup{
pdftitle={Unpacking ``Personal'' Health Informatics for Proactive Collective Care},
pdfsubject={cs.HC, cs.CY},
pdfauthor={Shyama Sastha~Krishnamoorthy Srinivasan, Mohan~Kumar, Pushpendra~Singh},
pdfkeywords={Ecology of tracking, Personal Health Management, Collective Care},
}

\begin{document}
\maketitle

\begin{abstract}
	Current mobile health platforms are predominantly individual-centric and lack the support for coordinated, auditable multi-actor workflows. However, in many settings worldwide, health decisions are enacted through multi-actor coordination rather than individual users. We present \textit{JEEVHITAA}, a cross-platform mobile system enabling role-aware sharing and verifiable information flows within permissioned care circles. \textit{JEEVHITAA} ingests platform and device data, builds layered profiles from sensors and tiered onboarding, and enforces fine-grained, time-bounded access control across care graphs. Data stays secure both within the application and the cloud. Integrated retrieval-augmented Large Language Models produce structured, role-targeted summaries and action plans, offer evidence-grounded verification with provenance and confidence scores, and support advanced insights on reports. We describe the architecture, connector abstractions, and security primitives, and report robustness evaluations using synthetic, ontology-driven data and findings from a feasibility study with real-life care circles across 9-14 weeks. We outline plans for larger multi-site evaluations focusing on operational alignment, longitudinal trust \& literacy impact, and relational friction \& efforts to sink into the daily infrastructure.
\end{abstract}

\keywords{Health Informatics \and Information Systems \and Collective Care}

\input{files/1_intro}
\input{files/2_related}
\input{files/3_design}
\input{files/4_implementation}
\input{files/5_method}
\input{files/6_eval_findings}
\input{files/7_discussion}
\input{files/8_conlusion}

\bibliographystyle{ACM-Reference-Format}
\bibliography{references}





\input{files/9_appendix}

\end{document}

%% file: files/1_intro.tex
\section{Introduction}

In many parts of the world, including large portions of India, care for everyday health and wellbeing is often collective \& cooperative. It remains a negotiated activity, distributed across networks of actors who share responsibility for detection, interpretation, and action. For instance, family members monitor symptoms, spouses/partners coordinate medication and appointments, friends/community health workers (CHWs) support during emergencies, and informal caregivers perform hands-on tasks as needed. These practices are shaped by household routines, availability rhythms, cultural expectations about privacy and authority, and negotiated interpretations of ambiguous signals. At the same time, consumer devices and mobile apps are rapidly expanding the range of measurable health signals (from vitals to activity, sleep, and rehabilitation exercises) and are being deployed to support recovery and long-term management. This growing instrumentation, however, is primarily designed for individual users. Notifications, interfaces, and decision support assume a single actor is responsible for interpreting data and taking action. The result is a mismatch between the social reality of collective care and the individualistic assumptions built into mainstream health technologies.

Seen from the human perspective, such a mismatch may cause concrete frictions: surveillance or visible artifacts that offend local norms; awkward, informal redistribution of access to devices; and additional emotional and coordination burdens on caregivers who already shoulder disproportionate care work. AI and algorithmic features, now increasingly embedded in wearable platforms and apps as recommendation engines, risk scorers, and personalized alerts, can amplify both harms and benefits. When AI outputs are tailored to a single assumed decision-maker, they may trigger misdirected alarms, create conflicting guidance across actors, or intensify negotiation work as families and care teams reconcile divergent recommendations. Conversely, AI also represents an opportunity. If systems could reason about responsibility over time, determine what information is appropriate to share with specific actors, and tailor how recommendations are routed and contextualized within household routines, algorithmic assistance could more effectively support distributed, collaborative care.

Empirical HCI research has repeatedly documented how technologies that impose an individual-centric frame of care produce these pitfalls in real deployments. Systematic reviews indicate that the vast majority of behavior change and health interventions remain strictly centered on the individual as a standalone entity, ignoring broader social dynamics \cite{10.1145/3706598.3714072}. This individualistic focus persists across everyday self-tracking architectures and micro-incentive systems designed to regulate wellness behaviors \cite{10.1145/3706598.3714208, 10.1145/3706598.3713650}, as well as tailored applications meant for managing specific clinical or physical conditions \cite{10.1145/3411764.3445695, 10.1145/3706598.3713224}. Furthermore, even as systems adopt sophisticated mechanics to personalize health strategies or adapt sensory feedback \cite{10.1145/2559206.2560474, 10.1145/3411764.3445619, 10.1145/3613904.3642651}, they inevitably frame interaction around a single body. This single-actor assumption extends into recent developments in explainable AI (XAI) and medical diagnostic interfaces, which prioritize optimizing metrics/explanations for an isolated user rather than collaborative reasoning and multi-actor interpretation \cite{10.1145/3544549.3585604, 10.1145/3706598.3714058, 10.1145/3613904.3642352}.

These dynamics are particularly evident in the Indian context, where health-tracking practices are often shared but not formally supported by technology. Children may monitor a parent's vitals, spouses coordinate adherence, and multiple household members collectively interpret information during an illness to determine next steps. Yet mainstream platforms typically provide no constructs to represent shifting caregiving roles, to encode household routines such as actor availability, or to negotiate the visibility of sensitive data based on context. Consequently, families routinely adopt fragile, improvised workarounds (sharing screenshots via messaging apps, verbal updates, or informal delegation of device credentials) to make systems `work' in situ. Such ad hoc practices underscore a practical shift from personal informatics to complex family and interpersonal informatics configurations in which care burdens are informally redistributed \cite{10.1145/2998181.2998362, 10.1145/3274396}. However, relying on these makeshift strategies leads to significant unintended consequences, compounding data-induced stress and introducing serious patient safety hazards within the care ecosystem \cite{Abdelaziz_Garfield_Neves_Lloyd_Norton_van_Dael_Wheeler_McLeod_Franklin_2024, 10.1145/3715336.3735746}. Ultimately, these workarounds undermine the very privacy and accountability properties needed for safe health technology, highlighting the critical need to break down strict ``personal'' boundaries to support proactive, collective care \cite{srinivasan2026unpackingpersonalhealthinformatics}.

From a systems perspective, these human-facing failures indicate absences in the technical abstractions that undergird current digital health ecosystems. The lack of role semantics, temporal access controls, and auditable cross-actor coordination leads to predictable system problems. We identify three technical limitations in current platforms that are particularly consequential for deployed, multi-actor settings. First, existing platforms lack representation of \textbf{collective responsibility:} they cannot explicitly encode who is responsible for a task, how responsibility shifts among actors, or how escalation should proceed if a primary actor is unavailable. Second, privacy controls are largely static and individualistic, unable to express \textbf{role-sensitive visibility, culturally grounded expectations,} or \textbf{time-bounded access} rules that reflect who may see what and when within a care circle. Third, AI-driven recommendations and alerts are delivered as if a single person is responsible for interpretation and action, whereas in practice, responses are often \textbf{collaborative, negotiated,} and \textbf{mediated by routines.} This leads to over- or under-action, conflicting decisions among actors, and additional emotional labor for those already bearing caregiving burdens.

To address these gaps, we propose \textit{JEEVHITAA} (Joint Engagement Ecosystem for Vital Health Insights, Tracking, and Actionable Assistance) as a socio-technical artifact designed to uncover if (and how) families can better negotiate privacy, trust, and care infrastructure. \textit{JEEVHITAA}'s design dimensions are organized around three principles:

\begin{itemize}
\item \textbf{Role-aware contextualization:} Encode care-circle rhythms, caregiving norms, and resource constraints into machine-readable policies and routing heuristics so alerts and interventions align with expected on-the-ground behavior.
\item \textbf{Fine-grained, auditable sharing:} Enforce role-sensitive, time-bounded access with efficient revocation, end-to-end encryption, and immutable provenance/audit logs to make information exchange accountable.
\item \textbf{Distributed responsibility \& actionable insights:} Provide a canonical responsibility model and actor-aware routing plus structured, provenance-tagged summaries so the right actor receives actionable information at the right time.
\end{itemize}

\textit{JEEVHITAA} treats the \textit{care circle} as the unit of design rather than positioning the individual as the sole locus of monitoring and action. Its implementation combines a cross-platform Flutter client with native Android integrations, platform and Bluetooth Low Energy (BLE) connectors, an encrypted local profile store (Keystore/Keychain), a versioned canonical ontology for profile items, a local/cloud vector index for retrieval, and a dual retrieval-augmented Large Language Model (LLM) pipeline that supports structured insight generation and automated verification. Key system capabilities include role-aware ACLs, provenance and audit logs for LLM outputs, revocation and key-rotation flows, and connector abstractions for vendor compatibility. We evaluated \textit{JEEVHITAA}'s technical consistency and its real-world impact. First, we stress-tested the system's functions using ontology-driven synthetic evaluations that measure summary fidelity, exposure control across different ACL configurations, provenance accuracy, and compatibility with mainstream wearable vendors. Next, we conducted a pilot deployment of the system to real-world care circles for a minimum duration of 9 weeks (max 14 weeks). The pilot study aimed at answering the following RQs, each addressing corresponding hypotheses:

\begin{itemize}
    \item Does the system effectively transition from ``User Management'' to ``Care Circle Support''?
    \begin{itemize}
        \item \textit{Hypothesis 1.1:} Multi-actor onboarding and device linking can be completed by diverse users (age 28–73) without continuous expert intervention.
        \item \textit{Hypothesis 1.2:} Users will report that role-based sharing controls (Subject vs. Care Actor) align with their ``on-the-ground'' care hierarchies and responsibilities.
    \end{itemize}
    \item How does the ``Neutral AI'' impact health literacy and information trust?
    \begin{itemize}
        \item \textit{Hypothesis 2.1:} The ``Verification Engine'' improves the care circle's ability to critically evaluate and filter external misinformation (e.g., WhatsApp shares).
        \item \textit{Hypothesis 2.2:} Explicit provenance badges and confidence scores increase the perceived reliability of health summaries compared to raw data feeds.
    \end{itemize}
    \item Does the system reduce the emotional and social ``cost'' of caregiving?
    \begin{itemize}
        \item \textit{Hypothesis 3.1:} Offloading verification to the AI ``Neutral Party'' reduces interpersonal Friction (e.g., nagging or ``policing'' medication) between care actors and the subject.
        \item \textit{Hypothesis 3.2:} Negotiated visibility settings and time-bounded access reduce the subject's feeling of surveillance (the ``panopticon effect'') while maintaining safety.
    \end{itemize}
\end{itemize}

Based on the analysis of pilot deployment data, we present our findings and outline future directions.

In this paper, we present (a) system design for collective care, embedded governance mechanisms, implementation architecture, and systemic evaluation of \textit{JEEVHITAA}; and (b) an in-the-wild evaluation using a pilot deployment of \textit{JEEVHITAA} to understand its effectiveness over time. We demonstrate how it fills persistent gaps in current digital health ecosystems and outline how its design principles enable robust, culturally legitimate, and collectively meaningful care coordination. By addressing both socio-technical and infrastructural challenges, particularly those revealed when AI is embedded in multi-actor settings, \textit{JEEVHITAA} provides a proof of concept for next-generation mobile health systems that work \textit{in conjunction with,} rather than \textit{in opposition to,} existing collective care structures.

%% file: files/2_related.tex
\section{Related Work}
Our work sits at the intersection of collective informatics, privacy as a negotiated practice, and AI-mediated health literacy. While prior systems research has focused on the feasibility of sensing, we focus here on the socio-technical orchestration of these capabilities.

\subsection{From Personal to Collective Informatics}
The dominant digital health paradigm, the ``stage-based model'' of Personal Informatics (PI) \cite{10.1145/1753326.1753409}, focuses on individual self-reflection and behavior change. However, HCI scholarship has long critiqued this individualistic focus, arguing that care is fundamentally social and distributed \cite{10.1145/3357236.3395469, 10.1145/3197391.3197394}\textemdash enacted within families, care networks, and other support circles rather than by isolated users \cite{10.1145/2998181.2998362}. Evidence from field studies like ``caring though data'' \cite{10.1145/2998181.2998303} and design frameworks such as ``distributed cognition'' and ``social coginition'' \cite{10.1145/353485.353487, 10.5555/AAI10223295} shows that tracking becomes part of caregiving practices, that families use shared dashboards and Do it yourself (DIY) infrastructures for remote monitoring, making it an emotional and relational practice, not just an instrumental one. Further studies \cite{10.1145/3544548.3581546, 10.1145/3715668.3736372} show that these sociotechnical practices create emotional labor, disclosure work, and negotiation over visibility and control\textemdash and further examine the importance of caring for caregivers separately \cite{10.1145/2441776.2441789}. Despite these observations and the theoretical shift, the technical architectures of mobile health remain stubbornly individual-centric. By treating the `user' as an atomic unit, they fail to support the `interpersonal informatics' required for coordination. Most `sharing' features in commercial apps are read-only dashboards that fail to support the dynamic handoffs and collaborative decision-making observed in the real world. \textit{JEEVHITAA} bridges this gap by implementing collective informatics not just as a User Interaction (UI) layer but as a foundational system architecture that treats the `Care Circle' as the atomic unit of design and explicitly models the distribution of responsibility.

\subsection{Infrastructuring and Coordination in the Wild}
Care is rarely enacted through a single app. It is `infrastructured' through a bricolage of tools, including WhatsApp groups, paper logs, and verbal handoffs. Works by Star \& Ruhleder \cite{Star_Ruhleder_1996}, Shove et al. \cite{shovesocial}, and Pipek \& Wulf \cite{Pipek_Wulf_2009} define infrastructure not as a `what' but a `when'\textemdash a system becomes infrastructure only when it sinks into the background of daily practice. In many collective societies worldwide, this `infrastructuring' relies heavily on social redundancy and improvisational coordination. Prior literature has explored such engagement through specific use cases, including dashboards for stress data \cite{10.1145/3563657.3596096}, supporting infrastructure for changing health goals \cite{10.1145/3715336.3735813}, wellbeing in the workplace \cite{10.1145/3610207}, family care coorination in pediatric care \cite{10.1145/3555187, 10.1145/3686998}, and even the use of defamiliarization in games to promote reflective thinking about health \cite{10.1145/3461778.3462046}. However, most current Health-AI systems assume a linear, clinical workflow (symptom $\rightarrow$ diagnosis $\rightarrow$ treatment) that clashes with these fluid realities. \textit{JEEVHITAA} supports these `assemblages' of care by integrating formal sensing (wearables) with informal coordination (interaction, delegation), allowing the system to adapt to the `rhythms' of care rather than imposing a clinical schedule.

\subsection{AI-Mediated Sensemaking and Actionability}
As health data becomes complex, the challenge shifts from access to interpretation. While recent advancements in Large Language Models (LLMs) show promise for summarizing medical data, `explainability' in a care circle is distinct from explainability for a single user. Amershi et al. \cite{10.1145/3290605.3300233}, and others \cite{10.1145/3656156.3665127} in Human-AI Interaction argue for systems that support human agency and collaborative sensemaking. In a care network, an `anomaly' detected by AI must be explained differently to a spouse (who needs to know ``Is this urgent?'') than to a clinician (who needs the specifics). Another part of sensemaking and actionability is how people receive health information/advice, and how they deal with it. Prior work has explored designing artifacts for health education \cite{10.1145/3715668.3735626}, data visualizations to support family members' understanding \cite{10.1145/3563657.3596109, 10.1145/3637334}, and qualitatively studying message-based dissemination systems \cite{10.1145/3643834.3661504}, among others \cite{10.1145/3706599.3720208}, for supporting health information understanding/education. However, current systems still do not explain the provenance of information (how the information is generated or where it is coming from). \textit{JEEVHITAA} contributes to this space by introducing provenance-aware AI, where the system not only summarizes data but tracks the ``chain of custody'' for insights, helping the care circle build trust in the automated guidance without displacing human judgment.

\subsection{The Gap: Operationalizing Collective Care}
Prior health system implementations provide sophisticated building blocks, yet they remain fundamentally optimized for the isolated `quantified self.' For instance, secure architectures like \textit{Amulet} \cite{10.1145/2676431.2676432} have successfully demonstrated how to manage low-power device-to-cloud synchronization for continuous biosensors, but their security frameworks are not designed around multi-user, role-based dynamics. Similarly, while software-defined platforms such as \textit{HealthSense} \cite{10.1145/3300061.3345433} offer powerful remote aggregation dashboards for managing tracking parameters over time, they lack the underlying system semantics needed to orchestrate or route tasks among informal domestic caregivers. Even recent analytical frameworks like \textit{GLOSS} \cite{10.1145/3749474}, which significantly advance open-ended sensemaking by using LLMs to triangulate passive wearable feeds, operate under the assumption of a single beneficiary, leaving the interpreted insights unaligned with a shared ontology.

This systemic anchoring to the individual limits the scope of otherwise advanced telemetry, rehabilitation, or ambient sensing designs in the literature (as detailed in Table \ref{appendix:mapping}). Without explicit multi-actor operational models, prior configurations suffer from three critical limitations:

\begin{itemize}
    \item First, systems remain \textit{atomistic}: they treat caregivers as `observers' or alert targets rather than active collaborators, lacking the stateful abstractions required to negotiate handoffs or shared responsibilities.
    \item Second, privacy controls remain \textit{binary and static}, failing to capture the `negotiated boundaries' of domestic life where access often depends on urgency and context \cite{10.1145/3715336.3735678} rather than permanent permission lists.
    \item Finally, coordination support is often \textit{institutional}, designed for clinical workflows rather than the informal, mixed-literacy networks at homes.
\end{itemize}

\textit{JEEVHITAA} bridges this gap not by advancing sensing accuracy, but by introducing new design abstractions for \textit{collective care}. We shift the unit of design from the individual to the care circle, operationalizing (a) role-aware access controls, (b) provenance-backed verification, and (c) time-bounded delegation. This approach transforms sensing from an end in itself into a substrate for durable, collaborative caregiving.

%% file: files/3_design.tex
\section{System Design and Architecture}
Drawing on Srinivasan et al.'s \cite{srinivasan2026unpackingpersonalhealthinformatics} qualitative work (including surveys, semi-structured interviews, and stakeholder interviews), which identified sharing as a central user need, guiding card-sorting-based co-design workshops with spousal pairs and interviews with family dyads, clarified what to share, with whom, and under what conditions. \textit{JEEVHITAA}'s initial design utilized these findings with a focused set of requirements: per-metric controls, action templates (monitor, nudge, emergency), approval flows, role-specific summaries, and confidence \& provenance indicators. This process helped us design not only the system but also its underlying technical architecture to encode collective care.

\begin{figure}[ht!]
    \centering
     \includegraphics[width=.99\linewidth]{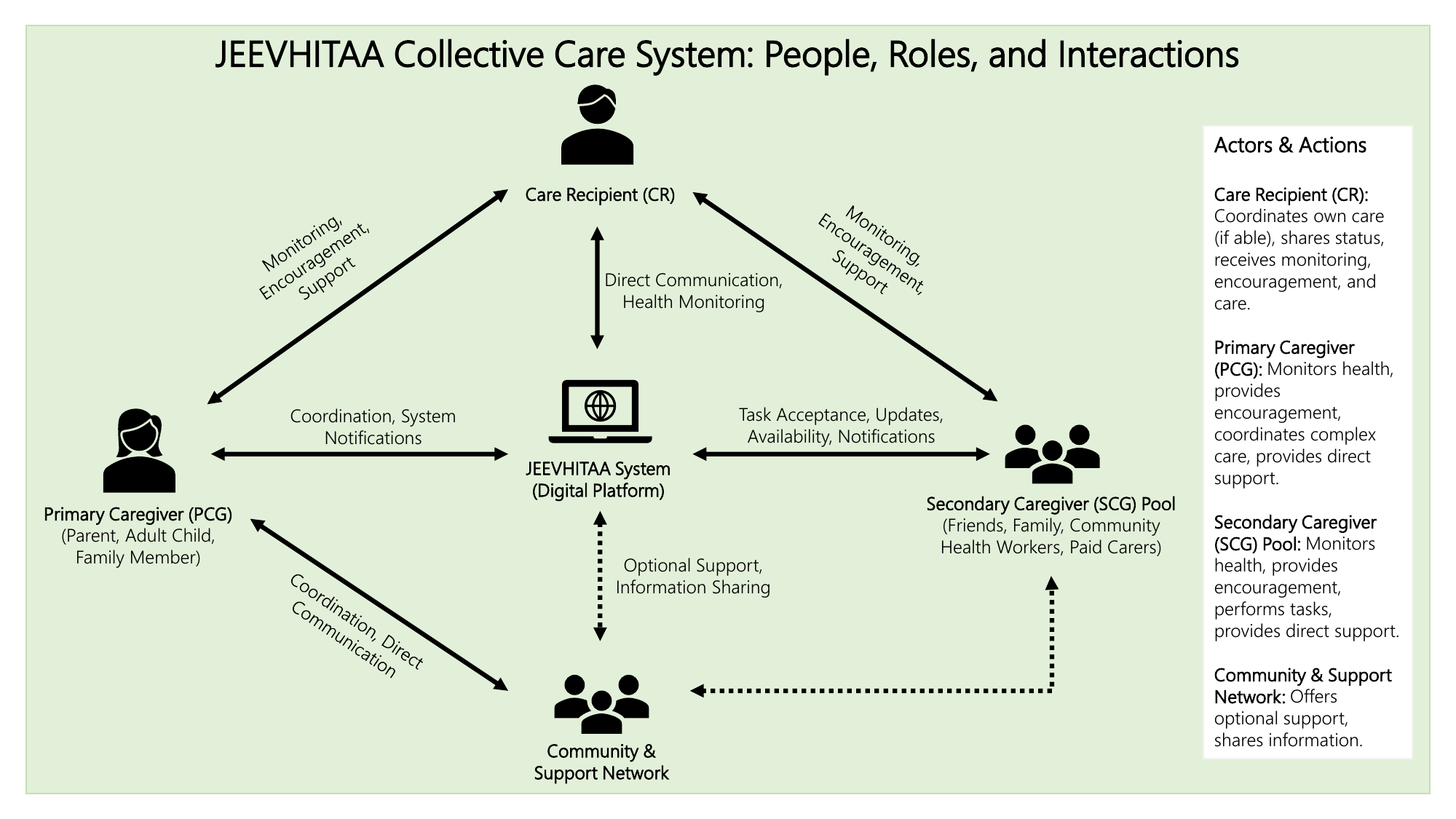}
     \caption{\label{Figure_if} Interactional flow of \textit{JEEVHITAA}, shown as a logical abstraction of distributed components.}
\end{figure}

\begin{figure}[ht!]
    \centering
     \includegraphics[width=.99\linewidth]{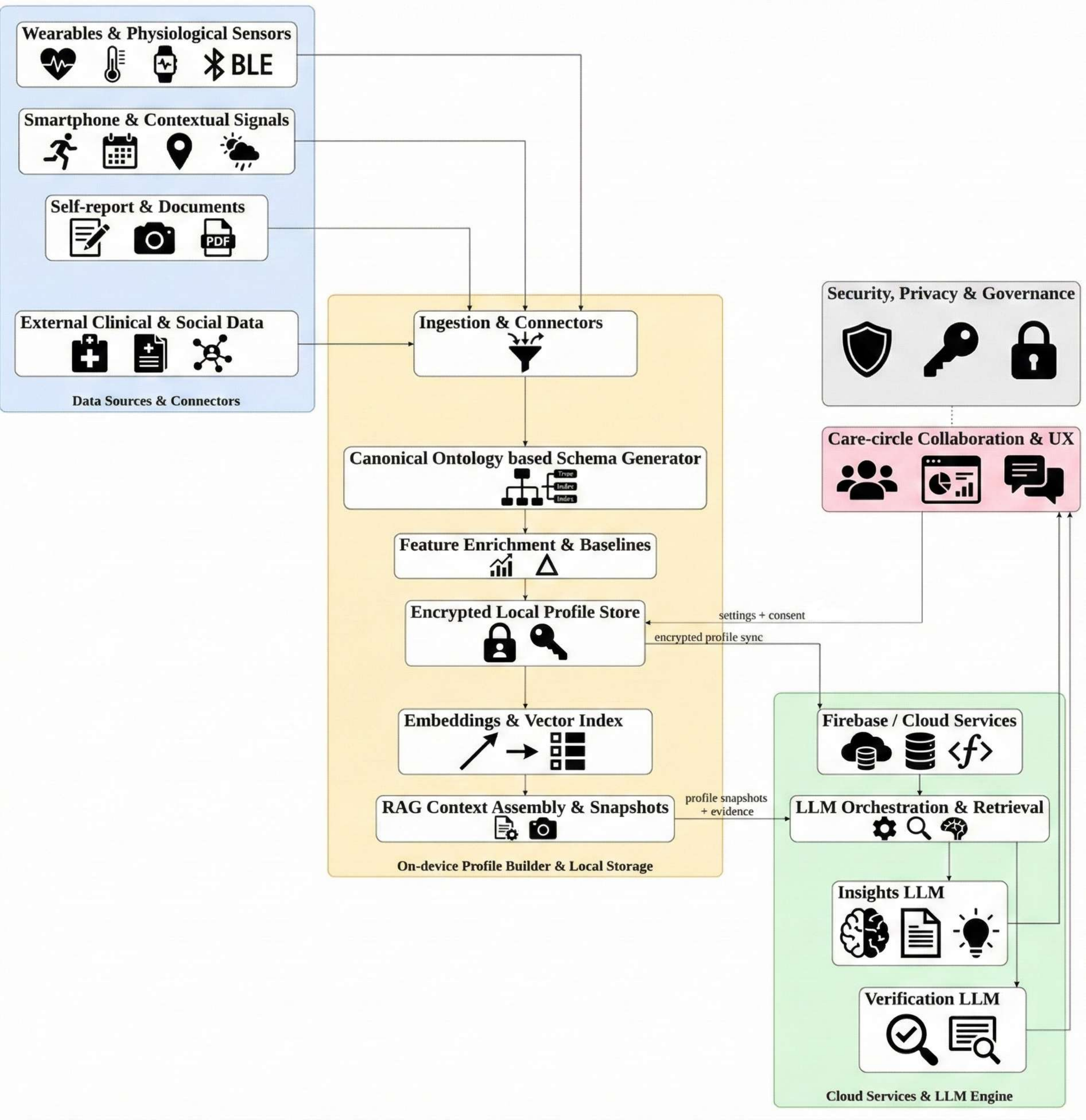}
     \caption{\label{Figure_arc} End-to-end architecture of \textit{JEEVHITAA}}
\end{figure}

\subsection{Iterative Design and Prototyping}
We adopted a participatory, iterative process to design and prototype \textit{JEEVHITAA}. Design proceeded in short, evidence-driven iterations. Themes were converted into interaction primitives and JSON card definitions, and medium-fidelity Figma screens were produced to realize a Data \& Sharing center, per-stakeholder templates, explicit approval flows, and stakeholder-specific export views. Each Figma build served as a probe in lightweight evaluations, during which participants provided think-aloud feedback and requests for plain-language explanations and export formats. Iteration reduced cognitive load and increased perceived control, yielding a small set of prioritized changes (clarified microcopy, confidence badges, and export templates) that were carried into specifications. Through these design outcomes, care circle dashboards were designed as the primary means for actors in a care circle to support the subject. Each care circle has a subject and 2 or more actors (including the subject) who can all view/interact with the subject's health summary and physiological metrics (depending on the access controls set by the subject for the actor's role).

Importantly, prototyping directly informed architectural decisions rather than remaining a visual exercise and design guide. Figure \ref{Figure_if} illustrates the interactional flow at the level of logical responsibilities, abstracting over deployment details. Requirement mappings and the Figma card schema made canonical data types and permission semantics explicit, which motivated the creation of a machine-readable ontology and a versioned profile schema to normalize sensor and health questionnaire inputs. Prototype approval templates and role definitions are translated into concrete access-control requirements, including consent records, per-metric ACLs, and provenance-based audit logs. Requests for stakeholder-specific exports influenced storage and sync formats, while the need for confidence indicators guided decisions about local preprocessing, provenance metadata, and the inclusion of a verification stage in the LLM pipeline.

The iterative cycle, therefore, fulfilled two complementary functions. It validated interaction designs with users, and it produced an executable specification that clarified system boundaries and data contracts. The final architecture (Figure \ref{Figure_arc}) maps directly to participant needs and prototype artifacts, and includes a canonical ontology \ref{appendix:ontology}, permissioned care-circle graphs, encrypted local stores with key management, connector interfaces for sensors and vendor Application Programming Interfaces (APIs), and a two-tier LLM pipeline for insights and verification operating over retrieval-augmented context. This traceable pipeline ensured that interface affordances such as agency and control were implemented as concrete architectural mechanisms.

\subsection{Architecture and Specifications}
The system is architected as an end-to-end platform that combines a cross-platform Flutter application with native platform extensions, cloud services, and an on-device local stack to provide secure, semantically coherent personal profiles and AI assistance. The client application is implemented in Dart, using native Android runners and a shared web target. This approach enables the core UI and business logic to be portable, while sensing and secure storage are implemented via platform channels and native plugins. Cloud services via Firebase provide authentication, cloud persistence, and serverless logic, enabling synchronized profile state, notifications, and server-side compute.

On the device, a local encrypted data store holds the canonical Profile and cached embeddings, with keys protected by Android Keystore. Background services ingest sensing data and orchestrate embedding and sync operations. A canonical ontology, provided as a \textit{TXT} file, defines entity types and relationships, serving as the lingua franca that maps heterogeneous sensor and self-report inputs into semantically typed profile entities. The AI layer distinguishes two LLM roles. An insights model Application Programming Interface (API) that generates personalized guidance and explanations, and a verification model (API) that fact-checks and cross-validates claims against structured profile data and trusted sources. Both utilize retrieval augmentation from a vector store of profile data embeddings.

Collaboration is enabled through a permissioned graph of users and care circles, with memberships, roles, and consent records enforced by cloud access control rules and mirrored locally for UI rendering. To contextualize the response/summary for each actor within the care circle, the LLM generates summaries along with role, membership, and action information specific to each actor. Security and privacy are built into every tier, with end-to-end encryption of Personally Identifiable Information (PII) in transit and at rest, minimization of exposure of raw sensor streams through local preprocessing, and audit logs that record LLM queries and shared insights for traceability.

%% file: files/4_implementation.tex
\section{Ecosystem Implementation}
The implemented ecosystem integrates modular client components, sensing connectors, an ontology-driven profile builder schema, storage with strong cryptographic controls, and an LLM orchestration layer that supports retrieval-augmented generation and verification. Connectors capture physiological, behavioral, and contextual signals and emit typed events that are validated and normalized by the ingestion pipeline. Normalized entities are persisted in an encrypted local store, indexed by a vector store for retrieval, and synced to the cloud as fine-grained consent and ACLs.

The AI orchestration layer composes compact profile snapshots and retrieved evidence to ground LLM responses, while a verifier model evaluates claims for contextual relevance and factual consistency. Care circles are instantiated from templated dashboards that render aggregated views and AI briefings while enforcing consent and role-based access. Developer extensibility is provided through well-defined connector interfaces, JSON card templates for dashboards, and configurable orchestration parameters for model endpoints, retrieval sizes, and verification thresholds. Finally, Continuous Integration (CI) checks validate ontology mappings and schema migrations to maintain compatibility across versions.

\subsection{Mobile Application}
Given the popularity and open-source nature, we chose Android as the initial implementation of our system, even though we built it using a cross-platform framework like Flutter. The Android target is implemented as a Flutter application shell hosted by the Android runner, with the presentation, domain, data, and services layers organized to separate the UI, business logic, persistence, and system integrations. Flutter widgets and screens render dashboards, care circle views, conversational AI interfaces, onboarding flows, and settings, and the UI adapts to platform styles. Background work is handled by native APIs exposed via platform channels: foreground services for continuous collection, WorkManager or an equivalent for scheduled jobs for periodic sync, and event callbacks to surface native events in Flutter.

Native code components provide advanced sensor access, including BLE and device-specific monitoring, Google Fit/Health Connect integration, encrypted keystore operations, and persistent background workers with permission management. The codebase follows a modular pattern, isolating presentation from domain use cases and data repositories. Dependency injection is used to manage service bindings, such as the LLM client, embedding engine, and encryption helpers. Runtime permission flows are implemented with just-in-time prompts during onboarding, and native notifications are employed to surface care circle alerts, urgent health flags, and consent changes to the user.

\subsubsection{Data Sources and Profile Builder}
Data sources include physiological sensors from wearables and BLE peripherals, smartphone inertial and contextual sensors, self-reported questionnaires and logs, passive contextual inputs such as sleep and calendar events, and social context such as care circle metadata. As part of the application's settings, the user can also set their app usage goals, step goals, daily time cycle (the duration of morning, afternoon, evening, and nighttime for their lifestyle), and the LLM Engine chosen for the insights generator and verifier. The profile builder functions as an ETL pipeline that ingests typed events with provenance and consent metadata, normalizes units and resolves outliers through smoothing and aggregation, and maps processed inputs to the canonical ontology.

Enrichment stages derive higher-level features, such as heart rate variability, activity classifications, and sleep quality scores. They compute baselines and deltas and persist normalized entities and timeline events in an encrypted local database, with optional cloud synchronization. Selected profile items, summaries, and important events are converted to embeddings and indexed in the local and cloud vector store to support retrieval-augmented reasoning. The profile schema is versioned, and migrations are applied on update to preserve compatibility. The ontology serves as the authoritative schema for entities, inclusive of but not limited to Person (individual), Vitals, Medication, Symptom, Event, Care Circle, and Consent Record.

\subsubsection{AI Assistance and Care Circle Dashboards}
The AI assistance component is designed to deliver personalized, evidence-based insights by combining carefully engineered prompt templates, concise profile summaries, and a small set of retrieved profile entries or guideline excerpts. System prompts define persona and privacy constraints; user prompts combine a succinct user query with structured context; and assistant prompts instruct the model to emit structured outputs that include summaries, observations, recommendations, confidence levels, and data references. Retrieval-augmented generation is implemented so that each dialog turn constructs a short profile snapshot and retrieves the most recently updated profile entries from the vector store. The model receives only this contextual window, thereby limiting exposure to the whole history. Responses are produced in a structured JSON-like format for deterministic parsing and also rendered as human-readable narratives in the conversation UI. All AI interactions are logged with metadata that records retrieved IDs and model signatures to support later verification and audit. To support users in ingesting external documents, an optional upload feature for PDFs, Docs, and images is provided as part of the AI-powered Report Analysis Widget. The same options are available in the contextually relevant information verification widget, along with videos, links, and forwarded messages.

\begin{figure}[ht!]
    \centering
     \includegraphics[width=.99\linewidth]{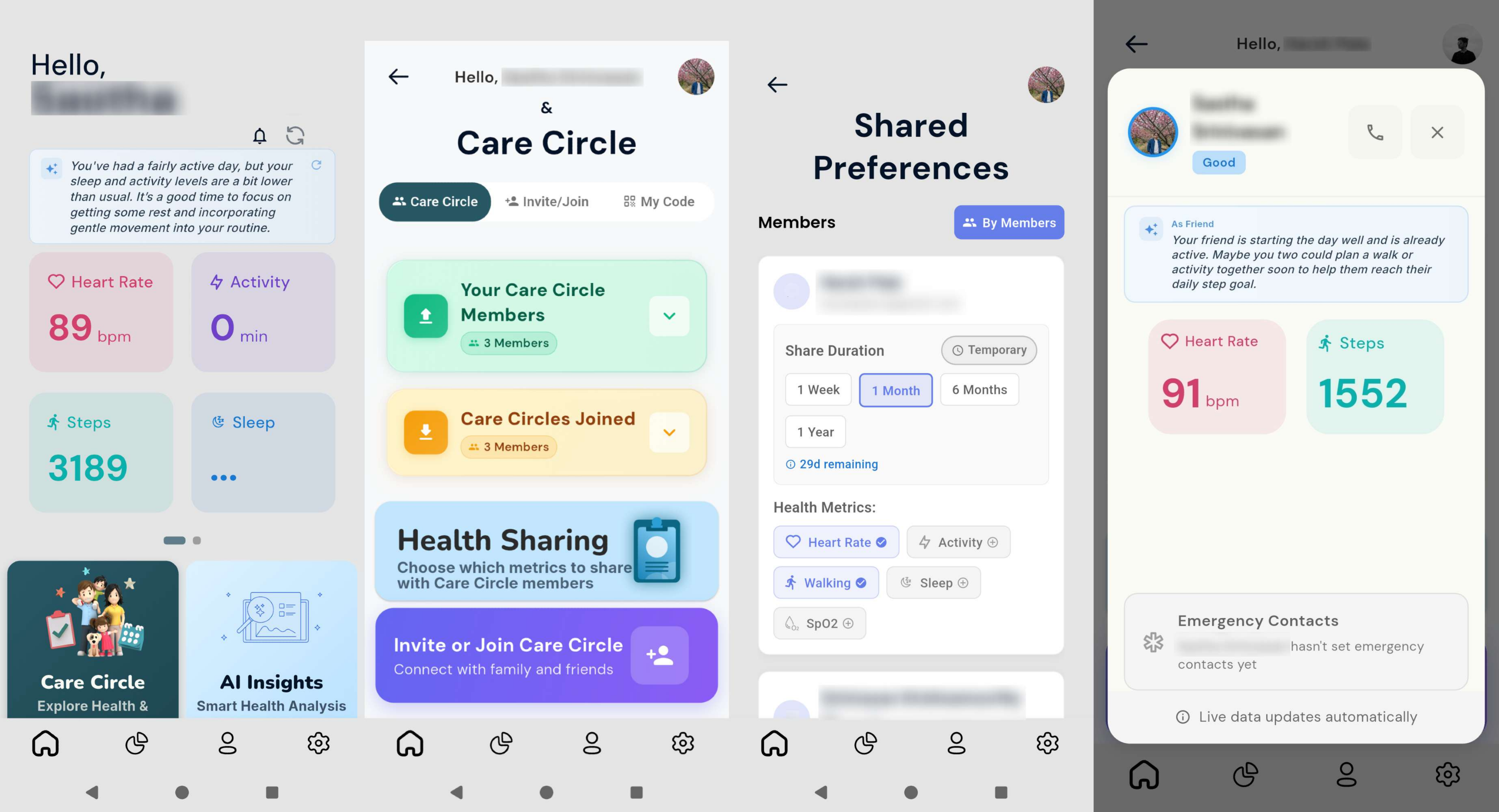}
     \caption{\label{Figure_ver} Android Interface - Profile, Sharing Controls Flow, Care Circles, and Shared Summary.}
\end{figure}

\begin{figure}[ht]
    \centering
     \includegraphics[width=.99\linewidth]{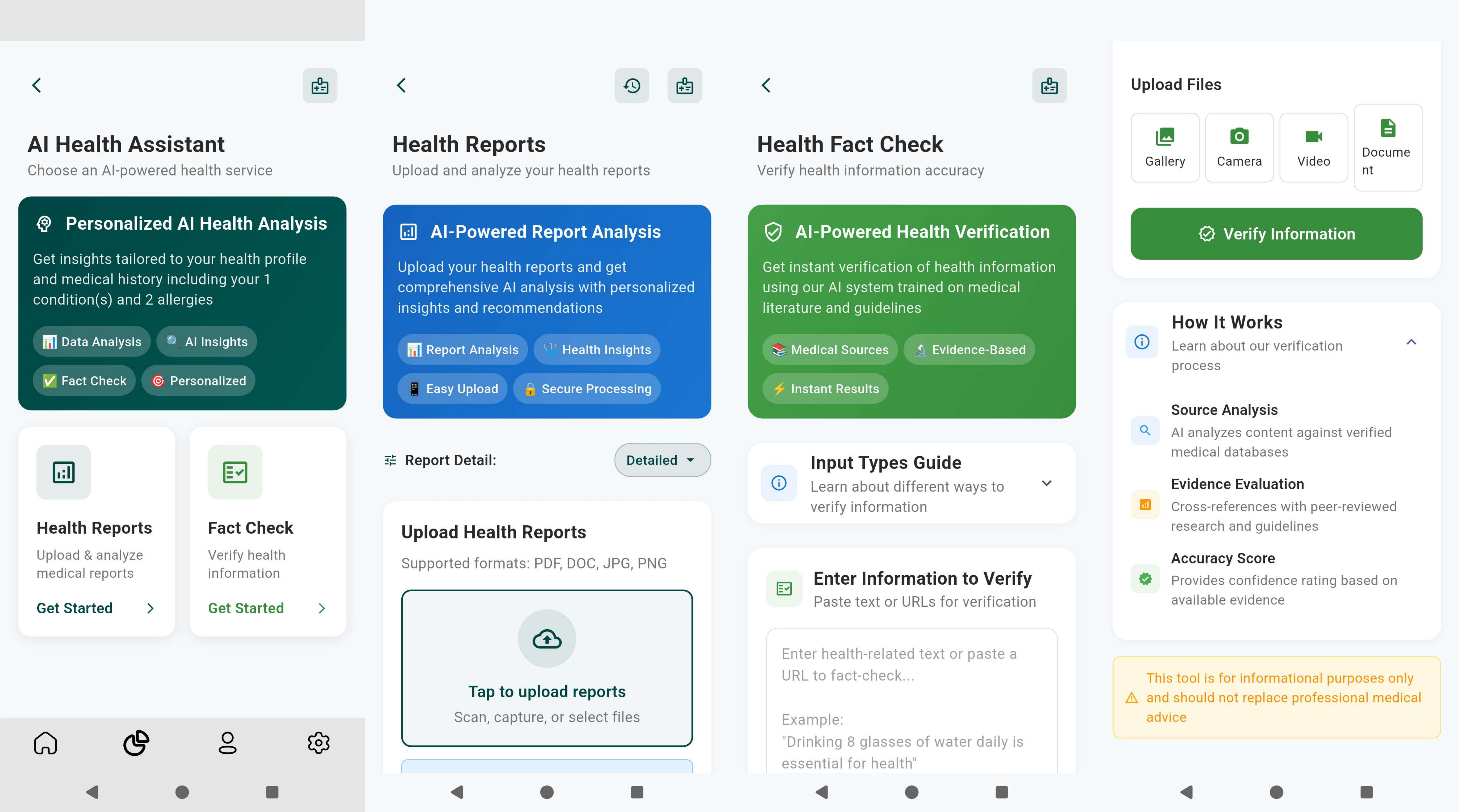}
     \caption{\label{Figure_ins} Android Interface - AI Assistance and Health Information Verification Engine.}
\end{figure}

\begin{figure}[ht!]
    \centering
     \includegraphics[width=.99\linewidth]{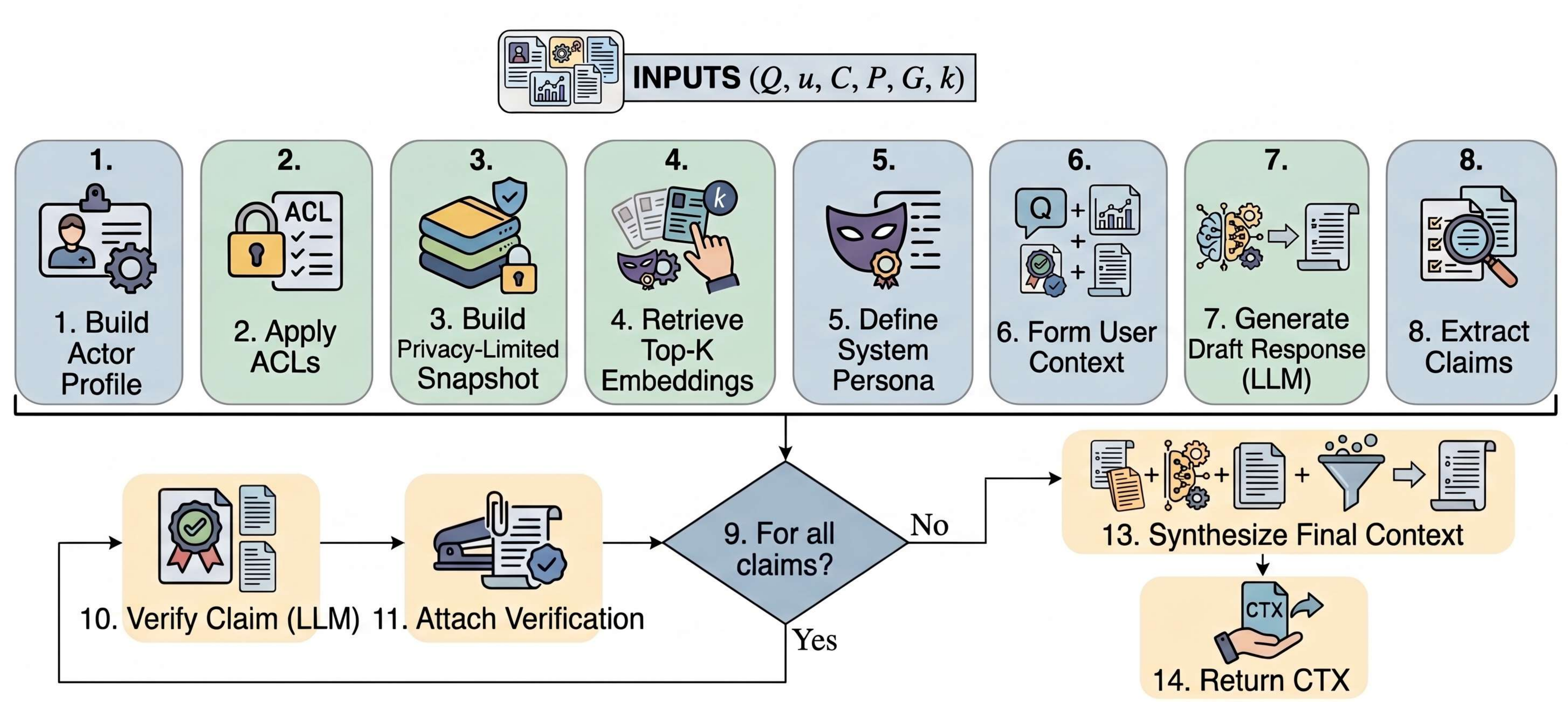}
     \caption{\label{Figure_verf} Algorithmic flow diagram of Verifer.\protect\footnotemark}
\end{figure}

\begin{figure}[ht!]
    \centering
     \includegraphics[width=.99\linewidth]{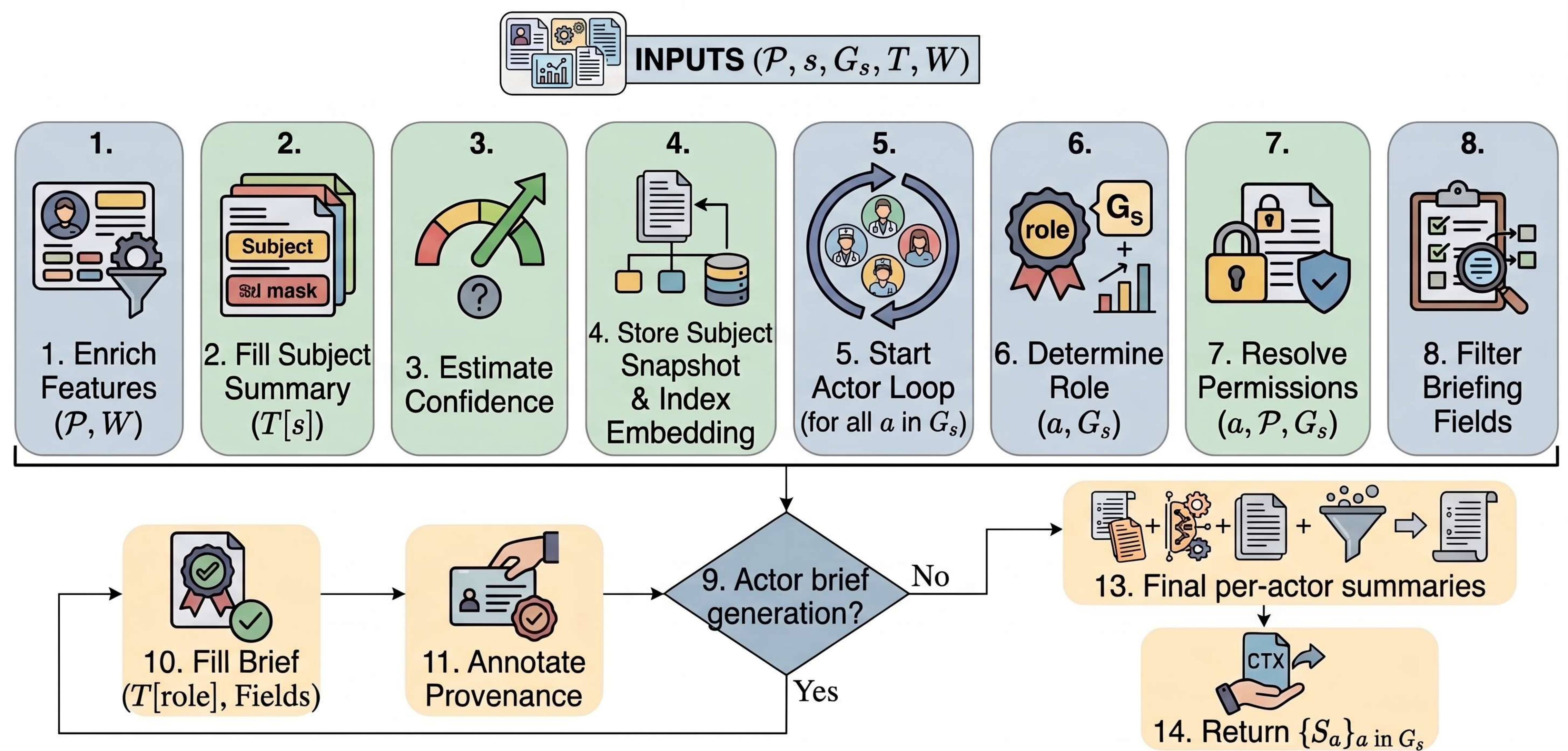}
     \caption{\label{Figure_insf} Algorithmic flow diagram of Summarizer.\protect\footnotemark[\value{footnote}]}
\end{figure}

\footnotetext{We thank Google's Gemini for assistance in the graphic visualization of the Algorithms.}

Care circle dashboards are constructed from reusable widget cards that display trends, alerts, recommendations, tasks, and shared notes. Templates for common monitoring scenarios are instantiated per circle, and card definitions are stored as JSON that reference ontology fields. Fine-grained permission and consent records determine which fields are visible to which roles, and an AI briefing operation composes a care-circle-scoped context, generating concise briefings or Q\&A responses that are constrained by access rules and roles. Dashboard creation follows a practical sequence in which permissions and consent flows are defined, JSON card schemas and widget libraries are implemented, server endpoints and access rules supply card data, the RAG pipeline is connected to the AI briefing button, and sharing, notification, and governance flows are added to enable revocation and auditability. Figure \ref{Figure_ver} illustrates a sample scenario in which a user enables the sharing of metrics with an actor from their care circle, and shows how the metrics and summary are then visible to the actor. When the user becomes part of someone else's care circle, the widget displaying their information also includes health status flags and a quick contact button that opens to phone/messaging options to contact the subject of the care circle directly, based on the user's choice. A widget for AI Insights leads to separate screens for both AI-powered Health analysis and contextualized health information verification (Figure \ref{Figure_ins}). All AI-assisted generation workflows include a collapsible widget that explains how it works and provides disclaimers about the need to consult experts/doctors before taking critical action based on the advice. The algorithms (presented in Figure \ref{Figure_verf} and \ref{Figure_insf}; detailed versions in Appendix \ref{appendix:algo}) represent the workflows of the health information verification engine and the profile summary generator, respectively.

\subsection{Sensing Connectors}
Sensing connectors conform to a common abstraction that exposes lifecycle methods, such as connect, authenticate, subscribe, sample, transform, and pushToStore, and they register metadata, including source type, sampling frequency, and consent requirements. Capture modalities include BLE GATT peripherals for heart rate and SpO2, Google Fit and HealthKit adapters for platform health data, vendor API adapters using OAuth for services such as Fitbit and Garmin, and on-device sensors, such as an accelerometer and microphone, for activity and optional voice features.

The capture pipeline buffers raw readings, performs noise filtering and unit conversion, detects spikes and extracts events, maps samples to ontology entities, and persists the resulting ProfileEvents in the encrypted local store while enqueueing items for cloud sync when permitted. Time alignment is managed by recording both device and server timestamps, along with timezone and offset metadata. Provenance records include source device identifiers, firmware versions, and sampling methods. Privacy-preserving defaults favor local preprocessing and storage of derived features rather than continuous raw streams, with opt-in controls for storing more sensitive traces.

\subsection{Storage and Encryption}
Storage is implemented across both local and cloud layers, with strong cryptographic protections and a key lifecycle that supports key revocation. Locally, structured profile data and events are persisted in an encrypted document store, such as SQLite with SQLCipher. Small items are stored in an encrypted key-value storage, and embeddings are retained in an encrypted local vector store. The symmetric keys that protect local stores are stored in the device keystore/secure enclave and are rewrapped by Firebase's cloud key management service for cross-device synchronization.

Firebase (cloud storage) uses managed document stores and object storage to sync profile state and larger artifacts. Sensitive fields are uploaded as ciphertext, preventing the server from decrypting them. TLS protects data in transit and may employ certificate pinning for critical endpoints. The encryption/decryption key is stored as an entity within the APK and cannot be retrieved without developer access. Audit logs are partitioned and stored to maintain tamper evidence and minimize exposure of raw personal data. Revocation flows will allow users to revoke cloud access, triggering key invalidation and, optionally, a local wipe of synced PII.

\subsection{LLM Engine}
The LLM engine is organized as a two-tier pipeline that separates generative insight from factual verification, reducing the risk of hallucination and increasing traceability. The insights model receives a structured prompt composed of a system persona, a short user query, and a compact profile snapshot, together with five to eight retrieved evidence snippets from the vector store. It is tasked with producing a structured draft comprising a summary, observations, recommendations, confidence levels, and referenced data IDs, all of which are contextual to the role in the care circle. Key claims from that draft are passed to a verification model, which evaluates each claim against the same retrieved items and curated guideline excerpts. The verifier then returns pass/fail decisions, along with evidence citations and confidence scores.

Retrieval augmentation is a central component of this workflow. Embeddings derived from profile entries, guideline snippets, and prior outputs are stored in a vector index, and the orchestration layer retrieves top-k items with similarity scores to form the contextual window. Prompting utilizes a fixed system prompt that defines constraints and output schemas, while runtime prompts concatenate retrieved context and user inputs. Context windows are deliberately small, allowing only the most relevant snippets to be transmitted. The orchestration pipeline may sequentially retrieve, generate, verify, and synthesize steps, and the engine supports model selection parameters that adjust creativity and determinism for the insights and verification roles, respectively. To preserve history without transmitting full chat logs, the system relies on compressed, periodically generated profile summaries, selective retrieval of top-k event embeddings, metadata pointers to profile item IDs, short conversational windows, and ephemeral per-turn context that is discarded after use; all LLM interactions record metadata such as prompt hashes, retrieved IDs, and model signatures for auditing. Operational safeguards include rate limiting, consent-bound invocation of LLM calls, deterministic fallback guidance when models are unavailable, and logging practices that record only hashed references to sensitive data while preserving the provenance necessary for later review.

%% file: files/5_method.tex
\begin{figure}[ht!]
    \centering
     \includegraphics[width=.99\linewidth]{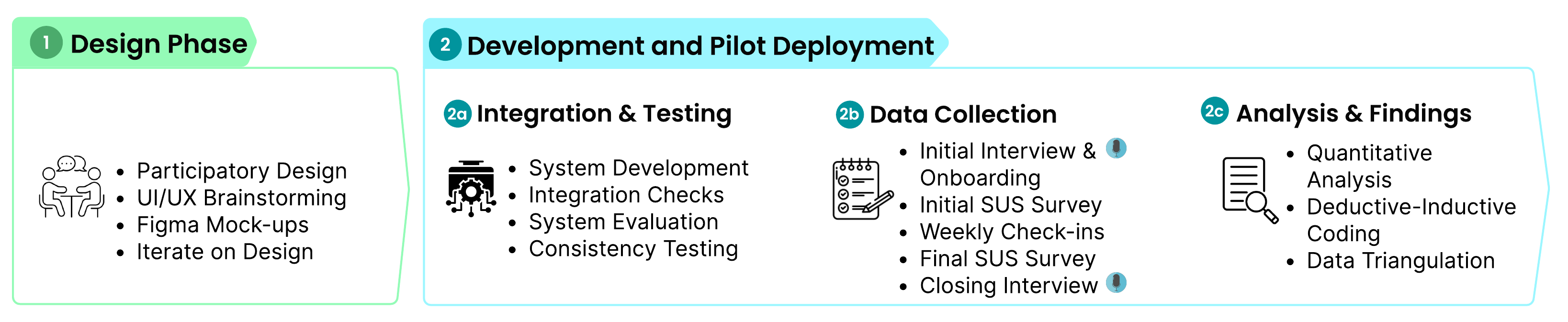}
     \caption{\label{Figure_jpwf} Study Workflow.}
\end{figure}

\section{Methods}
To evaluate the real-world utility, intergenerational dynamics, and the efficacy of the HCAI mechanisms embedded in JEEVHITAA, we conducted an in-the-wild, mixed-methods, longitudinal field deployment after approval from our institutional review board (IRB). Iterative prototyping for the system began in May 2025 and continued for 4 iterations, with participant feedback, to finalize the UI design by July 2025. After the prototyping workshops, we designed and developed the system as a fully functional support for collective care by October 2025 and conducted system testing through November 2025. The study transitioned from an initial short-term proof-of-concept (by the end of November 2025) to an extended evaluation, tracking five multi-actor care circles (by January 2026) over 9 to 14 weeks per circle.

\subsection{Study Design and Longitudinal Timeline}
The deployment was structured into five distinct chronological phases (presented in Figure~\ref{Figure_jpwf}) designed to capture both the baseline technical barriers and the evolutionary social dynamics of collaborative domestic caregiving:
\begin{enumerate}
    \item \textbf{Phase 1: Pre-Interviews \& Socio-Technical Baseline:} We conducted individual and group semi-structured interviews with all participants prior to onboarding. These interviews established pre-existing care configurations, manual communication workarounds (e.g., sharing medical reports or vitals via unstructured WhatsApp groups), domestic authority hierarchies, and individual privacy comfort thresholds.
    \item \textbf{Phase 2: Pre-SUS Assessment:} Immediately following system onboarding, platform configuration, and initial device/wearable linking, all participants independently completed a baseline System Usability Scale (SUS) survey to capture initial cognitive friction and onboarding usability benchmarks.
    \item \textbf{Phase 3: Active Longitudinal Deployment \& Weekly Check-ins:} Over the course of the 9-14 week deployment, participants used \textit{JEEVHITAA} naturally within their households. Continuous, automated background telemetry tracked device synchronization rates, access log entries, and queries executed via the AI Verification Engine. To prevent recall bias and capture micro-frictions, brief weekly qualitative check-ins were administered to track shifting social tensions, surveillance anxieties, and platform utility.
    \item \textbf{Phase 4: Post-SUS Assessment:} At the conclusion of each care circle's deployment window, participants re-administered the 10-item SUS instrument to evaluate longitudinal shifts in system learnability, perceived complexity, and operational acceptance.
    \item \textbf{Phase 5: Post-Interviews \& Provenance Quiz:} The evaluation concluded with extensive semi-structured group exit interviews. Embedded within these interviews was a hands-on, situational \textit{Verbal Comprehension and Provenance Quiz} to empirically verify whether the qualitative feelings of agency matched the users' actual technical comprehension of system data and privacy configurations.
\end{enumerate}

\begin{table*}[ht!]
\centering
\small
\begin{tabular}{llcllllcc}
\toprule
\textbf{Care} & \textbf{User} & \textbf{Age} & \textbf{Gender} & \textbf{Role within Circle} & \textbf{Highest} & \textbf{Tech} & \textbf{Pre} & \textbf{Post} \\
\textbf{Circle} & \textbf{ID} &  &  & & \textbf{Education} & \textbf{Proficiency} & \textbf{SUS} & \textbf{SUS}\\
\midrule
\textbf{JP1} (14) & JP1S  & 64 & Male   & Subject & Bachelor's & Intermediate & 80.0 & 97.5 \\
             & JP1A1 & 35 & Female & Actor (Daughter) & Bachelor's & Intermediate & 80.0 & 92.5 \\
             & JP1A2 & 31 & Male   & Actor (Son) & Master's & Advanced & 85.0 & 97.5 \\
             & JP1A3 & 28 & Female & Actor (Niece) & Master's & Advanced & 82.5 & 95.0 \\
\midrule
\textbf{JP2} (11) & JP2S  & 31 & Female & Subject (PCOS) & Bachelor's & Advanced & 87.5 & 92.5 \\
             & JP2A1 & 33 & Male   & Actor (Husband) & Bachelor's & Advanced & 85.0 & 97.5 \\
\midrule
\textbf{JP3} (11) & JP3S  & 29 & Female & Subject  & Bachelor's & Advanced & 85.0 & 97.5 \\
             & JP3A1 & 31 & Male   & Actor (Husband) & Bachelor's & Advanced & 85.0 & 95.0 \\
\midrule
\textbf{JP4} (11) & JP4S  & 38 & Female & Subject (Thyroid) & Bachelor's & Advanced & 80.0 & 92.5 \\
             & JP4A1 & 40 & Male   & Actor (Husband) & Bachelor's & Advanced & 82.5 & 95.0 \\
             & JP4A2 & 35 & Female & Actor (Sister-in-Law) & Bachelor's & Advanced & 82.5 & 92.5 \\
\midrule
\textbf{JP5} (9) & JP5S  & 69 & Female & Subject (Cardiac) & High School & Beginner & 82.5 & 90.0 \\
             & JP5A1 & 73 & Male   & Actor (Husband) & Bachelor's & Beginner & 77.5 & 87.5 \\
             & & & & (Asthma) & & & & \\
             & JP5A2 & 48 & Male   & Actor (Son) & Bachelor's & Intermediate & 85.0 & 95.0 \\
\bottomrule
\end{tabular}
\caption{Demographic Breakdown and Compositions of the Five Care Circles ($N=14$).}
\label{tab:participants}
\end{table*}

\subsection{Participants and Care Circle Compositions}
Participants were recruited through purposive and snowball sampling to capture diverse, multigenerational household structures typical of urban middle- to upper-middle socioeconomic strata in India (based on the updated Kuppuswamy scale by Saleem et al. \cite{saleemmodifiedkuppuswamysocioeconomic2021}). We primarily involved participants from traditional families in urban regions of Chennai, Bengaluru, and Delhi. The finalized study pool consists of five distinct care circles (\textit{JP1} through \textit{JP5}) comprising 14 participants (ages 28-73). To protect participant anonymity, all cohorts and individuals were assigned alphanumeric codes. Participants presented a wide spectrum of baseline technical proficiencies, ranging from \textit{Beginner} (requiring proxy assistance for setup) to \textit{Advanced} (highly independent tech adopters). Detailed demographics are presented in Table~\ref{tab:participants}; the minimum number of weeks in which all members of each circle participated is given in brackets next to the care circle code.

\subsection{Data Collection Instruments and System Infrastructure}
Our multi-channel data collection approach paired automated software telemetry with validated human-centric instruments:
\begin{itemize}
    \item \textbf{System Logs \& Telemetry Data:} The application platform enforced continuous server-side and local cryptographic logging. Telemetry pipelines recorded daily access/viewing rates for both the subject and the care actors in the care circle. The platform preserved immutable, timestamped audit entries that tracked every Access Control List (ACL) state mutation, every temporal permission grant, and every instance of manual access revocation executed by the subject. Additionally, all user queries routed to the \textit{AI Fact-Check/Verification Engine} were archived alongside system-generated confidence thresholds, prompt parameters, and factual provenance lineages.
    \item \textbf{Standardized Usability Scales:} The standardized 10-item System Usability Scale (SUS) instrument was administered individually at Phase 2 and Phase 4. This quantified the longitudinal shifts in system acceptability, capturing how extended, naturalistic exposure altered initial usability scores.
    \item \textbf{Semi-Structured Interviews and Diaries:} Phase 1 and Phase 5 interviews were conducted in native and preferred languages (English, Hindi, \& Tamil), audio-recorded, and fully transcribed. The weekly micro-check-ins captured during Phase 3 served as a running diary of localized friction and real-time negotiations within domestic spheres.
    \item \textbf{Provenance Comprehension Quiz:} Administered during the final group exit interviews, this contextual behavioral evaluation required subjects to actively navigate their live application interface to complete three verification tasks: (1) accurately identify the identity and exact timestamp of the last time a care actor viewed their physiological data, (2) actively modify or immediately revoke a care actor's temporal permissions, and (3) explain the source citation and logic behind an AI-generated ``Low Confidence'' notification flag.
\end{itemize}

\subsection{Trigger Events Observations}
To evaluate system routing resilience, role-aware visibility configurations, and collective cross-actor coordination patterns under stressful or anomalous contexts, we examined two possible trigger events within the first four weeks of the deployment:
\begin{itemize}
    \item \textbf{Trigger 1: Missed Care Event:} A simulated workflow anomaly where a critical daily check-in or medication logging prompt was intentionally ignored by the subject, triggering automated role-stratified escalation alerts across the care circle graph.
    \item \textbf{Trigger 2: Transient Elevated Heart-Rate Alert:} A physiological anomaly where a temporary, stationary elevated heart rate spike notifies the care circle and is questioned about it to see how care circle members utilize the provenance logs to audit data authenticity and trace the clinical lineage of the automated warning flag.
\end{itemize}

Using observations from these events, along with qualitative insights from the interviews, we effectively evaluated our system's routing resilience and its ability to understand role-aware visibility configurations.

\subsection{Data Analysis and Methodological Triangulation}
Quantitative analysis began by compiling and computing individual SUS response metrics from the baseline and post-deployment phases. Descriptive statistics mapped the longitudinal shifts across the cohort ($N=14$), yielding a baseline Pre-SUS mean score of $82.86$ ($SD = 2.75$, $\text{Min} = 77.5$, $\text{Max} = 87.5$) which significantly improved to a Post-SUS mean score of $94.11$ ($SD = 3.04$, $\text{Min} = 87.5$, $\text{Max} = 97.5$). This demonstrates a substantial positive delta ($\Delta = +11.25$) in usability over extended exposure. Background telemetry logs were processed to verify platform adherence, confirming that continuous data collection met our pre-defined feasibility threshold ($\ge 80\%$ active daily telemetry streams) across all deployed consumer hardware interfaces.

Qualitative transcripts from the baseline interviews, weekly check-ins, and final exit sessions were analyzed using a hybrid deductive-inductive thematic coding approach. Deductive codes were strictly anchored within our primary research questions (RQ1: Role-Based Alignment, RQ2: AI Sensemaking \& Provenance, RQ3: Relational Friction). Inductive codes were derived iteratively from emergent behavioral regularities observed during the multi-circle data analysis. The finalized codebook comprised four deductive codes (D) and five emergent inductive codes (I):
\begin{itemize}
    \item \texttt{ROLE\_ALIGN} (D): Mapping interface permissions and summary granularities directly to offline interpersonal roles (e.g., child vs. spouse).
    \item \texttt{AI\_VERIF} (D): Strategic interaction with the verification engine and confidence tags to evaluate external digital health misinformation.
    \item \texttt{FRIC\_RED} (D): Mitigation of active family tension by offloading monitoring reminders onto automated, system-driven push alerts.
    \item \texttt{PROV\_TRANS} (D): Perceived user sovereignty rooted in granular data provenance visibility and mutual audit trail tracking.
    \item \texttt{FLAG\_HEUR} (I): An emergent ``Traffic Light'' heuristic where care actors restrict deep text engagement unless an automated Status Flag changes color.
    \item \texttt{SOC\_SHIELD} (I): Utilizing neutral, system-generated summaries as an objective social excuse to initiate care interactions without sounding critical.
    \item \texttt{PROX\_AGNCY} (I): Intermediate users serving as configuration proxies for beginner subjects while utilizing transparency screens to preserve the subject's long-term digital independence.
    \item \texttt{CONT\_SKEP} (I): Contextual skepticism where advanced users verify AI claims directly against raw historical logs before extending system trust.
    \item \texttt{COLLAB\_MED} (I): A conceptual shift from a asymmetric ``Subject-as-patient'' hierarchy to a collaborative ``Care-Circle-as-Team'' dynamic.
\end{itemize}

Methodological triangulation was achieved by continuously cross-referencing subjective assertions from interview transcripts with objective, logged interactions recorded during the deployment. For instance, qualitative statements from subjects about improved personal agency and reduced surveillance anxiety were compared with telemetry data to assess how often they accessed their personal audit logs. These findings were then compared against the longitudinal usability improvements observed in the pre/post SUS metrics. Finally, participants' verbal claims of technical comprehension regarding AI outputs and privacy toggles were validated by their empirical performance on the Phase 5 Provenance Comprehension Quiz, ensuring that our qualitative conclusions are firmly grounded in verifiable system interactions.

%% file: files/6_eval_findings.tex
\section{Evaluations and Findings}
\label{sec:evaluations_and_findings}
Before launching our live field deployment, we subjected the \textit{JEEVHITAA} ecosystem to a rigorous, multi-tiered evaluation strategy to assess its algorithmic resilience and technical infrastructure. We first constructed an extensive synthetic corpus that accurately replicates the dense care structures and multi-modal health histories typical of Indian households. This dataset allowed us to evaluate the system's performance along two key operational vectors: the descriptive fidelity of role-filtered health summaries and the auditability of the underlying large language model (LLM) processing pipeline. Following these core tests, we conducted comprehensive cross-compatibility evaluations across diverse consumer wearable platforms to verify that our ingestion adapters could handle the real-world variety of sensors without fracturing user engagement. Finally, we deployed the fully integrated application into a naturalistic field environment through a pilot study to observe how these technical mechanisms adapt to the complex social dynamics of everyday collective caregiving.

\subsection{System Evaluations}
\subsubsection{Algorithmic Robustness and Indexing Performance}
To stress-test the system's data orchestration layers, we generated a synthetic corpus modeling $1,000$ unique care circles tailored to reflect urban and semi-urban household dependencies in India. These simulated circles had a median size of $4$ members, with an interquartile range of $3$ to $5$ actors, mirroring contemporary socioeconomic family trends. The dataset contained an average of $2,400$ telemetry samples per subject, with a controlled mean data-missingness rate of $12\%$ and an intentional injection of $3$ structural outliers per history to challenge the pipeline's preprocessing limits. Each profile trace was synthesized by sampling from the canonical ontology's entity fields to incorporate plausible demographic backgrounds, variations in vitals, clinical medication timelines, and device connectivity failures.

We evaluated the system under varied access control list (ACL) configurations, routing identical underlying subject histories through divergent permission levels to verify role-appropriate context isolation. Across $10,000$ simulated conversational turns, the text processing pipeline maintained exceptional accuracy. The core health insights generator achieved a precision of $0.94$, a recall of $0.89$, and an F1 score of $0.91$, while the dedicated verification pipeline achieved a precision of $0.92$, a recall of $0.88$, and an F1 score of $0.90$.

Our exposure control evaluations confirmed that the data shielding layer perfectly matched the configured boundary parameters. Automated field checks showed zero data leakage across predefined stakeholder viewpoints. The Care Primary role successfully exposed $15$ out of $15$ permitted fields, the Monitor role surfaced exactly $7$ out of $7$ authorized metrics, and the Nudge-only role accurately restricted visibility to its designated $3$ fields. Automated constraint checkers successfully detected latent policy conflicts in $2$ out of $120$ edge-case configuration templates, ensuring that rule contradictions could be flagged before deployment.

Furthermore, the verifier model routinely isolated inconsistencies introduced by the generative model, appending distinct evidence links back to local profile identifiers without exposing the broader historical data store. Localized baseline tracking and smoothing rules successfully mitigated false positives in actionable recommendations, validating the system's capacity to maintain strict data provenance and access boundaries within standard domestic network sizes.

\subsubsection{Wearable Integration and Resource Footprint}
To ensure our infrastructure could accommodate the diverse consumer hardware preferred by our target demographic, we evaluated the platform's connector abstraction layer against a broad suite of mainstream wearable ecosystems. This evaluation spanned platform SDKs, cloud APIs, and direct Bluetooth Low Energy (BLE) GATT profile parsers across major brands, including Google (Pixel and Fitbit), Samsung, Noise, Amazfit, Garmin, OnePlus, Xiaomi, Mobvoi, boAt, Oppo, and Noise. The end-to-end synchronization pass rates demonstrated excellent reliability across data ingestion modes. The Fitbit and Pixel configurations using Google Health Connect achieved a flawless $100\%$ integration pass rate; the Samsung Health native SDK adapters scored $98\%$; and Garmin's cloud OAuth framework recorded $94\%$. Deeper consumer device variants, such as Xiaomi and OnePlus architectures operating across mixed cloud-to-BLE environments, achieved a $88\%$ pass rate, while standalone lifestyle ecosystems, including Amazfit, boAt, and Noise, stabilized with pass rates ranging from $89\%$ to $95\%$.

Concurrently, continuous device profiling verified that the application's local stack did not compromise smartphone performance. When the application remained in an idle background state, average CPU consumption was limited to just $2\%$, rising to a modest $8\%$ during active local data ingestion workloads, and peaking at $18\%$ during active LLM inference turns. On-device memory allocation maintained a similarly tight footprint, using $45\text{MB}$ while idle, $120\text{MB}$ during active ingestion cycles, and peaking at $180\text{MB}$ during turn-based context synthesis. Hardware battery profiling conducted over a continuous 1-hour active operational simulation confirmed a marginal power drain of $0.8\%$ during idle background tracking, which scaled to a sustainable $4.5\%$ under active data ingestion workloads. These profiles demonstrate the long-term feasibility of maintaining persistent, secure background tracking on standard consumer mobile hardware. All stats are presented in the Table \ref{tab:evals}.

\begin{table*}[ht!]
\centering
\small
\begin{tabular}{ll}
\toprule
\textbf{Stat} & \textbf{Information} \\
\midrule
\textbf{Synthetic Data} & $1,000$ unique circles, median:$4$, IQR: $3-5$, Average samples/subject: $24,000$,\\
& Mean data-missingness: $12\%$, Intentional structural outlier: $3$\\
\midrule
\textbf{Generator} & Precision:$0.94$, Recall:$0.89$, F1:$0.91$ \\
\midrule
\textbf{Verifier} & Precision:$0.92$, Recall:$0.88$, F1:$0.90$ \\
\midrule
\textbf{ACL Testing} & Exposure: $100\%$ for all roles ($15/15$, $7/7$, $3/3$), policy check: $2/120$\\
& (policy updates as expected based on injection)  \\
\midrule
\textbf{Integration} & Health Connect ($100\%$), Samsung SDK ($98\%$), Garmin cloud Auth ($94\%$),\\
& mixed cloud-to-BLE ($88\%$), standalone ecosystems ($89\%$-$95\%$) \\
\midrule
\textbf{Performance} & CPU - idle: $2\%$, injest: $8\%$, active: $18\%$ \\
& On-device memory - idle: $45\text{MB}$, active: $120\text{MB}$, peak: $180\text{MB}$\\
& Battery (1-hour sim) - idle: $0.8\%$, active: $4.5\%$\\
\bottomrule
\end{tabular}
\caption{Information on the systemic evaluations.}
\label{tab:evals}
\end{table*}

\subsection{Pilot Study Findings}
To understand how these underlying technical mechanisms reshape the human realities of collaborative health tracking, we transitioned from algorithmic simulations to an in-the-wild field deployment. Our deployment tracked five distinct multi-actor care circles ($JP1$ through $JP5$), each comprising $14$ active participants spanning ages $28$ to $73$. Over an intensive longitudinal observation window of $9$ to $14$ weeks per household, these families integrated \textit{JEEVHITAA} into their standard daily routines. Following the deployment protocols detailed in the previous section, we introduced two workflow challenges during the initial month to stress-test the system: a \textit{Missed Care Event} (Trigger 1) and a \textit{Transient Elevated Heart-Rate Alert} (Trigger 2). The resulting empirical data form a rich narrative of how collective health technologies live, adapt, and settle in the domestic sphere.

\subsubsection{From Rigid Tracking to Shared Awareness: Nurturing Family Dynamics}
In the wild, the transition from conventional, individualistic monitoring tools to our collaborative care circle system reshaped how families communicated about daily health markers. Rather than enforcing uniform visibility, the system's role-aware summaries naturally adjusted the data volume to match each individual's relational boundaries. In care circle $JP1$, the subject ($JP1S$, 64M) noted that the platform effectively filtered out routine behavioral data `noise' for his family, which immediately changed their interaction style. His daughter and niece stopped demanding granular updates on his micro-activities and instead focused on his macro-level energy trends. This shift significantly reduced his sense of being handled, while allowing his daughter ($JP1A1$, 35F) to preserve her core family identity. She reflected on this balance in her exit interview:

\begin{quote}
``It gave me a high-level vibe, and I did not need to be a doctor; I just needed to be a daughter (laughs).'' - $JP1A1$
\end{quote}

This preservation of relationship boundaries over raw metric checking was mirrored across other households. For younger partners balancing chronic conditions, like the couple in circle $JP2$, the automated summary model eliminated the recurring mental load of verbal explanations. The subject ($JP2S$, 31F) emphasized that the app handled the daily repetitive chore of explaining her physical exhaustion due to Polycystic Ovary Syndrome (PCOS), transforming her husband's interactions from diagnostic inquiry into supportive alignment: ``It shifted us from `Why are you tired?' to `I see you had a rough day, let’s rest.''' Her husband ($JP2A1$, 33M) highlighted how this dynamic governed his behavior, explaining that he relied primarily on quick visual status flags rather than dissecting raw biometric values.

This interaction pattern revealed an essential decision-making shortcut, which we formalized in the codebook under the inductive label \textit{FLAG\_HEUR} (Status Flag Heuristic). Caregivers across multiple circles routinely adopted a `traffic light' approach, deliberately avoiding dense health records or AI text summaries unless a color-coded status indicator explicitly shifted away from green or provided a text label that warranted intervention.

Quantitatively, this integration was reflected in a dramatic upward shift in systemic trust and ease of use, evident by the increased SUS scores at the end of the study. Across the entire deployment cohort ($N=14$), the average System Usability Scale score rose from an initial pre-deployment baseline of $82.86$ ($SD=2.75$) to an optimal post-deployment average of $94.11$ ($SD=3.04$), representing a substantial positive delta ($\Delta = +11.25$) as the software integrated into their households, providing the descriptive context that triangulates and supports the qualitative themes.

Crucially, the field data also highlighted how family structures adapt to technical barriers. While advanced users navigated the settings independently, beginner- and intermediate-level users required collective support. In care circle $JP5$, the subject ($JP5S$, 69F) relied on her intermediate-proficiency son ($JP5A2$, 48M) to manage her initial onboarding, wearable authentication, and permission periods. This cooperative configuration model, inductively coded as \textit{PROX\_AGNCY} (Proxy Agency), showed that low-literacy technical layers do not compromise user sovereignty. Instead, older subjects maintained active control over their health spaces by utilizing the platform's retrospective view logs to review exactly how and when their configuration proxies adjusted permissions. In the context of traditional Indian multigenerational households, where hierarchical respect and parental authority are paramount, this mechanism allows elders to accept essential digital labor from younger family members without relinquishing their domestic sovereignty. The audit log effectively realigns the digital asymmetry to match the home's existing cultural grammar.

\subsubsection{The Objective Mediator: AI Sensemaking and the De-escalation of Domestic Friction}
A primary challenge identified by families during pre-deployment interviews was the emotional cost of tracking, often described as the exhausting transition of family members into the role of health police. Participants' experiences revealed that \textit{JEEVHITAA}'s dual-stage verification and insights architecture dramatically lowered this interpersonal friction by acting as a neutral, third-party buffer. By transferring the task of alerting and reminding to an objective system, the app absorbed the social tension traditionally caused by tracking, thereby enabling collective sensemaking.

In circle $JP2$, the subject ($JP2S$) noted that receiving an automated step or medication reminder directly on her mobile interface felt like an objective tool working as intended, whereas the same reminder issued verbally by her husband was frequently perceived as intrusive management. Her husband ($JP2A1$) verified this shift, stating that reviewing the dashboard silently confirmed her activity goals and allowed him to step back from monitoring her routines, removing a long-standing point of friction from their evenings. Similarly, the son in circle $JP5$ ($JP5A2$) reported that checking the daily status flag allowed him to stop calling his mother with defensive checking questions (e.g., ``Did you take your pill? Did you walk?''), converting their conversations from an administrative lecture into an active, supportive relationship.

Concurrently, this neutral AI layer altered how multi-generational households processed external digital health misinformation, a dynamic tracked under the deductive code \textit{AI\_VERIF}. During the study, participants frequently routed unverified health claims, predominantly ungrounded medical tips forwarded in family chat groups, through the Verification Engine. In care circle $JP1$, the subject ($JP1S$) uploaded a viral social media post regarding an unverified ayurvedic remedy for hypertension. The system returned a definitive `Low Confidence' badge ($15\%$ score), explicitly noting the complete absence of clinical evidence or baseline support. $JP1S$ chose to ignore the tip, prioritizing the digital badge over the strong social preference of his peer group. His son ($JP1A2$) reflected on the interpersonal utility of this mechanism: 

\begin{quote} 
``Having the verifier to show `this is unverified' with proof saved us a lot of bickering... I think he trusted the system's `badge' more than my opinion because he could see the medical sources linked right there.'' - $JP1A2$
\end{quote}

This pattern underscores how the AI verifier serves as a crucial sociocultural buffer in close-knit networks, where directly rejecting or questioning a health tip forwarded by a peer or family elder is often viewed as disrespectful. By offloading skepticism onto an objective, provenance-backed technical artifact, the system provides a polite, face-saving `social shield' for declining ungrounded practices without disrupting relational harmony.

Rather than fostering isolation, these system-generated summaries frequently functioned as a collaborative resource, coded inductively as \textit{SOC\_SHIELD} (Social Shield). When the system identified a multi-day decline in sleep efficiency or step counts, family members used the text as an objective opening for a conversation. In circle $JP1$, an automated sleep warning prompted the niece ($JP1A3$) to lead a lighthearted family discussion about whether the subject's restlessness stemmed from a new mattress or his late-night tea habits, turning a sensitive health concern into fun banter. In circle $JP4$, the sister-in-law ($JP4A2$) used an automated low-energy flag to justify visiting the home and assisting with grocery shopping and household tasks, thereby reducing the feeling of being a burden by the subject. This shift, noted under the code \textit{COLLAB\_MED}, indicates that structured AI summaries can successfully promote a shift from unequal oversight relationships to collective care provision.

\subsubsection{Bidirectional Accountability: Redefining Privacy through Reciprocal Transparency}
A central concern when deploying continuous physiological tracking within domestic spaces is the panopticon effect. The uncomfortable sense of being constantly watched by one's family. However, the field deployment revealed that subjects did not experience heightened surveillance anxieties. This comfort was directly achieved through the system's commitment to bidirectional transparency and real-time audit logs, a mechanism categorized under the code \textit{PROV\_TRANS}. By providing subjects with clear visibility into the chain of custody of their data, the system balanced the power dynamic between the observer and the observed.

In circle $JP1$, the subject ($JP1S$) emphasized that knowing he could review exactly who accessed his metrics, and at what precise moment, fundamentally altered his experience of the technology. Reviewing his activity logs revealed that his daughter only checked his active heart rate trends a few times a week, transforming his perception of the app from an invasive surveillance instrument into a supportive resource:

\begin{quote}
``Knowing `who is looking at what and when' makes it easier to feel in control. I was not being surveilled; I was being supported, and I had the logs to prove it.'' - $JP1S$
\end{quote}

This architecture proved equally powerful for advanced users like the subject in circle $JP3$ ($JP3S$), who noted that the ability to easily toggle temporary permissions or execute a total data blackout over holiday weekends gave her a strong sense of personal boundaries. Her husband ($JP3A1$) agreed, explaining that clear mutual visibility discouraged unnecessary snooping and fostered mutual respect for privacy within the household.

This qualitative sense of digital control was confirmed by the objective outcomes of our Phase 5 Verbal Comprehension and Provenance Quiz. Administered during final exit interviews, the quiz required subjects to independently perform complex privacy adjustments on their live application interfaces. Subjects across all circles successfully navigated their active dashboards. Intermediate and advanced subjects ($JP1S$, $JP2S$, $JP3S$, $JP4S$) recorded a $100\%$ task completion rate, successfully identifying the exact timestamp of their last caregiver access event, correctly explaining the data sources underlying their sleep metrics, and modifying granular metric toggles without assistance.

Even within circle $JP5$, the beginner-proficiency mother ($JP5S$) successfully interpreted the color-coded health status markers and demonstrated how to initiate person-based sharing durations, though she required physical steering from her son to isolate specific sub-menus. These findings demonstrate that privacy in collective informatics does not require complete data isolation; rather, it is maintained by providing users with clear data provenance, mutual visibility, and explicit control over their shared environments.

%% file: files/7_discussion.tex
\section{Discussion}
\label{sec:discussion}

\subsection{Operationalizing `Care' in Code: Beyond Atomic User Abstractions}
A primary contribution of our field evaluation with \textit{JEEVHITAA} lies in demonstrating how digital health platforms can transition from individual-centric systems to structures that mirror collective care graphs. Traditional mobile applications treat the user as an atomic, isolated unit, managing security through rigid, binary access controls that fail to reflect the dynamic nature of domestic spaces \cite{10.1145/1753326.1753421}. As Helms and Fernaeus \cite{10.1145/3461778.3462025} note, domestic care is fundamentally fluid, requiring constant negotiation, shared accountability, and impromptu support. By formalizing the care circle as the baseline architecture, \textit{JEEVHITAA} bridges the persistent `socio-technical gap' outlined by Ackerman \cite{10.1207/S15327051HCI1523_5}, aligning programmatic access rules with the lived realities of households.

Our long-term observations support Hypothesis 1.2, demonstrating that role-based context configuration (\textit{ROLE\_ALIGN}) successfully isolates semantic details within the boundaries of individual relations. Rather than overloading every family member with continuous streams of raw biological signals, the architecture isolates high-level indicators for distant observers while routing actionable metrics to primary caregivers. This stratification prevents data inflation, which typically causes fatigue in family structures \cite{10.1145/2998181.2998362}. 

However, the field deployment also revealed an important nuance in independent system configuration across diverse age groups. While Hypothesis 1.1 suggested that all family members would navigate onboarding uniformly, our empirical observations in care circle $JP5$ revealed the need for a collective onboarding strategy. Older subjects, operating at beginner proficiency levels, initially required technical scaffolding from family to execute platform permissions and pair hardware. 

Crucially, this dynamic did not compromise user sovereignty. Instead, it surfaced a pattern we term Proxy Agency (\textit{PROX\_AGNCY}). Even when technical execution was delegated to an intermediate family proxy, older subjects maintained high structural oversight by utilizing the platform's transparent, retrospective view logs to audit their proxy's configurations. This cooperative engagement model expands upon classic personal informatics models \cite{10.1145/1753326.1753409}, showing that usability in collective ecosystems is not merely an individual attribute but a shared family capacity. This collective adaptability is quantitatively reinforced by the significant positive delta ($\Delta = +11.25$) in SUS scores across the entire cohort, proving that sustained engagement with role-stratified interfaces builds confidence over time across varying literacies.

\subsection{AI as a Social Buffer: Mediating Truth and Preserving Relationships}
In contemporary health informatics literature, automated verification engines and large language models are almost exclusively evaluated through the lens of standalone accuracy or anomaly detection. Our empirical findings suggest a broader socio-technical role for AI within collaborative tracking environments: the model acts as a vital mediator of trust and interpersonal de-escalation. Interpersonal tracking within families often introduces significant emotional labor, where necessary care monitoring (e.g., confirming medication adherence or checking activity baselines) is frequently perceived as intrusive policing or nagging by the recipient.

By transferring the task of alerting and evaluation to an objective system, \textit{JEEVHITAA} alters these interpersonal communication dynamics. This transition supports Hypothesis 3.1, demonstrating that offloading monitoring logic onto automated pushes mitigates domestic friction (\textit{FRIC\_RED}). The application absorbs the social cost of tracking; notifications are interpreted as a neutral tool working as configured rather than a family member exercising authority. This buffer allowed caregivers to step back from manual policing, transforming interactions from defensive checking routines into supportive relationships. This operational behavior embodies what Bogdanova \cite{10.1145/3656156.3665127} terms the `aesthetics of algorithmic care,' in which automated systems function as active, neutral intermediaries that preserve the underlying social fabric among human users \cite{10.1145/3313831.3376590, 10.1145/3757599}.

Concurrently, this neutral position enhanced family health literacy by providing an authoritative shield against external digital health misinformation, supporting Hypothesis 2.1 (\textit{AI\_VERIF}). When processed through the Verification Engine, ungrounded social media tips were checked against local profile data and peer-reviewed literature. The presence of clear, explicit confidence badges and data-source links enabled subjects to quickly evaluate external claims, often choosing to dismiss unverified peer guidance despite strong social conformity. 

This sensemaking model introduces two important patterns of human-AI interaction. 
\begin{itemize}
    \item First, advanced users frequently demonstrated Contextual Skepticism (\textit{CONT\_SKEP}), checking primary clinical sources before accepting automated guidance and using the AI as a search assistant rather than an uncritical oracle \cite{10.1007/978-3-030-60117-1_33}.
    \item Second, caregivers heavily relied on a Status Flag Heuristic (\textit{FLAG\_HEUR}) to make traffic-light decisions, treating color-coded indicators as shortcuts. This practice demonstrates that the value of human-centered explainability in multi-actor spaces is not tied to data abundance, but to highly scannable, grounded summaries that protect caregivers from cognitive overload.
\end{itemize}

\subsection{The Burden of Visibility: Shifting from Surveillance to Symbiosis}
A persistent barrier to the adoption of continuous physiological sensing within domestic environments is the panopticon effect\textemdash the uncomfortable sense of constant, asymmetric surveillance experienced by the tracking subject \cite{10.1145/2998181.2998303}. Increasing data visibility to ensure safety often shifts the burden of household labor, introducing new pressures to perform `good health' for the family network \cite{10.1145/3544548.3581546}. \textit{JEEVHITAA} directly engages with this tension by implementing strict bidirectional transparency and granular, time-bounded consent features.

Our field results support Hypothesis 3.2, showing that explicit data provenance and clear access tracking help counteract surveillance anxieties (\textit{PROV\_TRANS}). When subjects can easily review exactly who accessed their vitals, and at what precise timestamp, the power asymmetry between the observer and the observed is balanced. Knowing that a daughter or spouse reviews only detailed metrics during automated alert states helps transform the user's perception of tracking from invasive surveillance to a supportive safety net. This alignment respects the core requirements of Contextual Integrity \cite{Nissenbaum_2004}, ensures that tracking boundaries match domestic social norms, and protects the subject's personal dignity \cite{10.1145/3272973.3273012}.

Crucially, this dynamic challenges design assumptions that fundamentally treat privacy as an individual fortress meant to enforce total data isolation. Within a collective informatics framework, domestic privacy is inherently negotiated and relational. Trust is not preserved by locking family members out, but rather by enforcing strict, reciprocal visibility that balances the power dynamic between the observer and the observed. Our findings show that structured AI summaries do not isolate family members; instead, they function as an objective conversational catalyst, or Social Shield (\textit{SOC\_SHIELD}). When the system flagged multi-day physical declines, family members used the automated text as a non-threatening reason to initiate support. This mechanism allowed families to organize collaborative actions\textemdash such as scheduling group walks or redistributing household chores\textemdash without making the subject feel criticized. This behavioral evolution, formalized under our codebook as \textit{COLLAB\_MED}, indicates that collective care platforms can successfully transition families from asymmetric monitoring hierarchies into symmetric, collaborative care teams.

\subsection{Limitations and Future Work}
While our field deployment successfully validated the care circle model's functional mechanics, several limitations limit the generalizability of our findings. First, our deployment cohort was limited to five care circles ($14$ participants) located in urban households with middle- and upper-middle-class socioeconomic status in India. While this demographic accurately reflects early adopters of consumer wearables, it implicitly centers on close-knit family structures. Collective care takes diverse structural forms globally, including chosen families, decentralized community networks, and professional caregiver arrangements. Future iterations could explore how the role abstractions embedded in \textit{JEEVHITAA} adapt to more fluid, non-traditional social structures. Second, we only tested an Android variant of our system. Given interest from iOS users as well, the next version would deploy a secondary pilot with both Android and iOS users, adding the ability to view content on the web via a QR code scan for utility. This will help us understand how these access configurations withstand the friction of cross-ecosystem support and long-term domestic changes. Finally, due to the pilot nature of the system, key-rotation procedures were not utilized. We plan to include those, rewrap keys, and migrate encrypted artifacts as needed for the next iteration.

To address these limitations, we are planning a phased, multi-site longitudinal evaluation that tracks diverse care networks over extended observation periods (spanning 4, 8, 12, and 16 weeks). This upcoming study will explicitly investigate the maintenance and repair work required as the platform settles into long-term domestic routines. We will automatically collect encrypted, provenance-rich telemetry and responsibility-state transitions, governed by machine-readable JSON policy packages. We intend to triangulate these automated metrics with structured qualitative outcome targets, specifically measuring the operational alignment between system-modeled rules and actual human care tracking (targeting $\approx\geq 80\%$ timing alignment, $\approx\geq 90\%$ responsibility agreement, and $\approx\geq 80\%$ provenance comprehension on exit quizzes). By releasing anonymized deployment artifacts and verified configuration templates, we aim to provide a clear benchmark for replication for the broader human-computer interaction community exploring secure infrastructure in collaborative health spaces.

\subsection{Synthesizing Core Design Principles}
Taken together, our empirical findings illustrate how \textit{JEEVHITAA} operationalizes the three core design principles established at the outset of this work. 
\begin{itemize}
    \item First, \textit{Role-Aware Contextualization} is realized through role-stratified summaries that filter out daily data noise, protecting individual relationship boundaries and honoring family hierarchies.
    \item Second, \textit{Fine-Grained, Auditable Sharing} is achieved through explicit data provenance and bidirectional access logs, thereby balancing the domestic power dynamic and mitigating surveillance anxieties.
    \item Third, \textit{Distributed Responsibility and Actionable Insights} are supported by the neutral AI verification pipeline, which effectively filters health misinformation, accounts for the social costs of alerting, and provides a collaborative opening for family care coordination.
\end{itemize}

By embedding these capabilities directly within the technical architecture, \textit{JEEVHITAA} demonstrates a path forward for mobile health platforms that support, rather than disrupt, the social reality of collective care.

%% file: files/8_conlusion.tex
\section{Conclusion}
In this work, we introduce \textit{JEEVHITAA}, a complete HCAI system designed to support collective caregiving, treating the care circle rather than the individual as the unit of design. It integrates a canonical ontology, permissioned care graphs, encrypted multimodal profiles, and a dual-LLM pipeline for health insights and health-tip verification. As the primary artifact contribution, we present the implemented system, along with systemic evaluations using synthetic data, compatibility tests for integrating various health wearables, and a pilot study to verify the system's usability in real-world care circles. In doing so, we substantiate our claim that supporting collective care requires infrastructural support for role-aware context, accountable sharing, and distributed responsibility; not merely individual-facing AI features. As the results show promise, planned phased longitudinal deployments in the wild will measure coordination, responsibility alignment, provenance comprehension, and system performance, while enforcing ethical safeguards and supporting the release of anonymized artifacts for future reproducibility.

%% file: files/9_appendix.tex
\appendix

\section{Usability Questionnaire}
\label{appendix:sus}

The following System Usability Scale (SUS) was administered to the Subject and Care Actors to evaluate the usability of the JEEVHITAA system. Participants rated each item on a 5-point Likert scale ranging from ``Strongly Disagree'' (1) to ``Strongly Agree'' (5).

\begin{enumerate}
    \item I think that I would like to use JEEVHITAA frequently.
    \item I found JEEVHITAA unnecessarily complex.
    \item I thought JEEVHITAA was easy to use.
    \item I think I would need support from a technical person to use JEEVHITAA.
    \item I found the various functions in JEEVHITAA (e.g., dashboards, AI verification) were well integrated.
    \item I thought there was too much inconsistency in this system.
    \item I would imagine that most people would learn to use JEEVHITAA very quickly.
    \item I found JEEVHITAA very cumbersome to use.
    \item I felt very confident using JEEVHITAA.
    \item I needed to learn a lot of things before I could get going with this system.
\end{enumerate}

\section{Interview Guide Pre-Deployment}
\label{appendix:interviewguidepre}

This semi-structured interview guide focuses on capturing the socio-technical baseline of existing care practices, daily life rhythms, and interpersonal power dynamics.

\begin{itemize}
    \item \textbf{Care Circle \& Roles:} Think about the last time you needed help with a health task. Whom did you ask? What was their role? How did you decide they were the right person to help?
    \item \textbf{Daily Rhythms (Cultural Grounding):} Are there specific times in your day (like afternoon rest or evening walks) when receiving a health reminder would feel like an interruption rather than help?
    \item \textbf{The `Nagging' Factor:}
    \begin{itemize}
        \item \textbf{For the Subject:} Does it ever feel like your family is `policing' your health?
        \item \textbf{For the Caregiver:} How do you feel when you have to remind the subject about medication/activity?
    \end{itemize}
    \item \textbf{Information Sources:} Where do you usually get health advice (e.g., WhatsApp, TV, neighbors)? How do you decide if a `health tip' you received on your phone is actually safe to follow?
    \item \textbf{Care Circle Dynamics \& Hierarchy:}
    \begin{itemize}
        \item \textbf{Task Distribution:} Can you walk me through a typical week of health-related activities? Who usually tracks the medication, who checks the vitals, and who decides when to see a doctor?
        \item \textbf{Decision Authority:} If there is a disagreement about a health decision (like whether to follow a new diet or take a specific pill), how is it resolved? Who has the final say?
        \item \textbf{Role Perception:} How do you describe your ``role'' in the household's health? Do you feel like a leader, a helper, or the person being looked after?
        \item \textbf{Information Flow:} When someone gets a health report or a lab result, who is told first? Is there any information you deliberately keep to yourself? Why?
        \item \textbf{Visibility Comfort:} On a scale of ``not at all'' to ``completely,'' how comfortable are you with your [Family Member Name] seeing your daily activity levels or heart rate? Any specific concerns?
    \end{itemize}
\end{itemize}

\section{Interview Guide Post-Deployment}
\label{appendix:interviewguidepost}

The Post-deployment interview was split into two parts. The first part focuses on triangulating system logs with the qualitative experience of the HCAI features, and the second part aimed to verify the effectiveness of transparency primitives by focusing on provenance, accuracy, and interpretability.

\subsection{Core System Observations}

\begin{itemize}
    \item \textbf{On Roles:} When you assigned the [Son/Daughter] role, did the summary tailored for them change in a way that matched what you wanted them to know?
    \item \textbf{On Verification:} You used the AI Fact Check for a WhatsApp tip. Did the `confidence score' make you feel more sure about ignoring or following the advice?
    \item \textbf{On Friction:} Did you feel the app `nagged' you less than your family members used to, or did it feel like just another person watching you?
    \item \textbf{On Provenance:} When the system flagged [Trigger Event X], was it clear where that data came from (e.g., the watch vs. your manual log)?
    \item \textbf{The `Panopticon' Check:} Did you feel more `watched' because the app was tracking you, or did the audit logs (knowing who looked at what) make you feel more in control of your privacy?
    \item \textbf{Collaborative Action:} Describe a time the care circle acted together based on a `Daily Briefing.' Who initiated the conversation, and how did the app's summary help that discussion?
\end{itemize}

\subsection{Verbal Comprehension \& Provenance Quiz}
\begin{itemize}
    \item \textit{Part A: Provenance \& Accuracy}
    \begin{itemize}
        \item \textbf{The `Who' Question:} Looking at your Access logs, can you tell me the last time [Care Actor Name] looked at your heart rate data? Were you able to understand why they looked at it?
        \item \textbf{The `Source' Question:} In your health summary, it says your sleep was `Average.' What does that mean for you? Can you show me where the app says it got that information? Is it from your watch, your manual log, or both?
    \end{itemize}
    
    \item \textit{Part B: Interpretability}
    \begin{itemize}
        \item \textbf{AI Fact-Check Attribution:} Point to a specific claim made in the last AI briefing. Can you show me the `Source' link it used to justify that claim? If you clicked that link, what would you expect to see?
        \item \textbf{The Confidence Check:} If the verifier gives a health tip a `Low Confidence' score but provides three citations, would you trust it? Why or why not?
        \item \textbf{The Access Check:} You granted the [Care Actor Name] access for 24 hours. If they try to look at your reports tomorrow afternoon, what do you think the app will show them?
        \item \textbf{Authority Levels:} As the `Subject,' can you show me the button to `Revoke' someone's access immediately? If you pressed it now, how would they know their access had been revoked?
    \end{itemize}
\end{itemize}

\section{Prior Systems for Health and Wellbeing}
\begingroup

\hyphenpenalty=10000
\exhyphenpenalty=10000
\label{appendix:mapping}
\begin{table*}[ht!]
    \centering
    \RaggedRight
    \begin{tabular}{@{}p{0.36\columnwidth} p{0.10\columnwidth} p{0.16\columnwidth}
                    p{0.16\columnwidth} p{0.12\columnwidth}}
    \toprule
    \textbf{System} & \textbf{Primary Focus} & \textbf{Physiological Monitoring} & \textbf{Remote Access} & \textbf{Multi-actor Coordination} \\
    \midrule

    AutoSense (\cite{10.1145/2070942.2071027}) &
    Stress &
    Yes, biosignals &
    Dashboards &
    No \\

    \addlinespace[2pt] 

    Amulet (\cite{10.1145/2676431.2676432}) &
    Platform &
    Yes, biosensors &
    device$\leftrightarrow$cloud &
    No \\

    \addlinespace[2pt] 

    Bioimpedence (\cite{10.1145/2594368.2594369}) &
    Bio ID &
    Yes, body sensors &
    No, research &
    No \\

    \addlinespace[2pt] 

    Ultra-low power (\cite{10.1145/2801694.2801710}) &
    Sensing &
    Yes, sampling &
    No &
    No \\

    \addlinespace[2pt] 

    pH Watch (\cite{10.1145/3307334.3328583}) &
    Sensing &
    Yes, PPG-derived &
    telemetry &
    No \\

    \addlinespace[2pt] 

    HealthSense (\cite{10.1145/3300061.3345433}) &
    Trials &
    Yes, aggregation &
    Yes, Dashboard &
    No \\

    \addlinespace[2pt] 

    Eating Detect (\cite{10.1145/3307334.3328565}) &
    Eating &
    Yes, behavioral &
    No &
    No \\

    \addlinespace[2pt] 

    RehabPhone (\cite{10.1145/3386901.3389028}) &
    Rehab &
    No, task-based &
    Yes, clinician &
    No \\

    \addlinespace[2pt] 

    Painometry (\cite{10.1145/3386901.3389022}) &
    Pain &
    Yes, biosignals &
    Yes, dashboard &
    No \\

    \addlinespace[2pt] 

    NeckSense (\cite{10.1145/3397313}) &
    Eating &
    Yes, multimodal &
    No &
    No \\

    \addlinespace[2pt] 

    In-ear thermos (\cite{10.1145/3384419.3430442}) &
    Core Temp &
    Yes, temperature &
    No &
    No \\

    \addlinespace[2pt] 

    Fatigue (\cite{10.1145/3460421.3480429}) &
    Fatigue &
    Yes, physiological &
    Yes, alerts &
    No \\

    \addlinespace[2pt] 

    MyDJ (\cite{10.1145/3491102.3502041}) &
    Eating &
    Yes, chewing &
    No &
    No \\

    \addlinespace[2pt] 

    PROS (\cite{10.1145/3495243.3560533}) &
    Energy &
    Yes, efficient &
    No &
    No \\

    \addlinespace[2pt] 

    APG (\cite{10.1145/3570361.3613281}) &
    Cardiac &
    Yes, cardiac &
    Yes, telemetry &
    No \\

    \addlinespace[2pt] 

    Asclepius (\cite{10.1145/3636534.3649366}) &
    Auscultation &
    Yes, PCG &
    Yes, televisit &
    No \\

    \addlinespace[2pt] 

    OpenEarable 2.0 (\cite{10.1145/3712069}) &
    Sensing &
    Yes &
    Yes, dashboard &
    No \\

    \addlinespace[2pt] 

    GLOSS (\cite{10.1145/3749474}) &
    Sensemaking &
    Yes, wearable &
    Yes, dashboard &
    No \\

    \bottomrule
    \end{tabular}
    \caption{Representative HCI, Sensing, and UbiComp systems and why they do not enable collective care (multi-actor). Each entry lists single-word feature summaries (Yes/No + qualifier (if relevant)).}
\end{table*}

\endgroup

\section{High-level Ontology for an Individual’s Ecosystem Mapping in Collective Care}
\label{appendix:ontology}
This appendix presents the ontology used in the design blueprint. It specifies the entities, attributes, and relationships that define the ecosystem for collective care. The ontology of a \emph{Person} is structured into three main layers: \emph{Intrinsic Factors}, \emph{Extrinsic Factors}, and the \emph{Care \& Interaction Factors}. Each layer expands on the respective attributes in greater detail.

\subsection*{Person}
person\_id: Unique identifier for the individual (UUID or ABHA ID or stable ID).

\subsection{Intrinsic Factors}

\subsubsection{Demographics}
\begin{description}
  \item[user\_id] Linkage to ecosystem record for backward compatibility.
  \item[age] Numeric value (in years).
  \item[birthDate] ISO 8601 date of birth.
  \item[gender] Male, Female, Other, Prefer not to say.
  \item[sex\_assigned\_at\_birth] Male, Female, Intersex, Unknown.
  \item[marital\_status] Single, Married, Divorced, Widowed, Cohabiting.
  \item[education\_level] None, Primary, Secondary, Higher, Vocational.
  \item[occupation] Free-text or controlled vocabulary.
  \item[employment\_status] Employed, Unemployed, Student, Retired.
  \item[economic\_status] Income quintile or SES scale.
  \item[region] Geographic location, region code.
  \item[urbanicity] Rural, Semi-urban, Urban.
  \item[religion] Free-text or controlled vocabulary.
  \item[community] Ethnic or cultural group.
  \item[languages\_spoken] List of ISO language codes.
  \item[literacy\_level] None, Basic, Intermediate, Advanced.
  \item[identity\_notes] Notes on self-identification or special categories.
  \item[legal\_status] Citizen, Refugee, Migrant, Stateless.
  \item[contact\_info] Phone, address, email.
\end{description}

\subsubsection{Physiology}
\paragraph{Anthropometrics}
\begin{description}
  \item[height] Centimeters.
  \item[weight] Kilograms.
  \item[BMI] Body Mass Index.
  \item[waist\_circumference] Centimeters.
  \item[body\_composition] Fat percentage, lean mass, water distribution.
\end{description}

\paragraph{Vitals}
\begin{description}
  \item[resting heart rate] Beats per minute.
  \item[blood pressure] Systolic/diastolic.
  \item[respiratory rate] Breaths per minute.
  \item[oxygen saturation] Percentage.
  \item[sleep hours] Average hours per day.
  \item[sleep quality] Self-reported, scale 1–5.
  \item[activity summary] Steps, activity minutes, sedentary hours.
\end{description}

\paragraph{Laboratory Values}
\begin{description}
  \item[HbA1c] Percent.
  \item[Hemoglobin] g/dL.
  \item[Vitamin D] ng/mL.
  \item[Lipid panel] HDL, LDL, triglycerides, total cholesterol.
  \item[Renal function] Creatinine, eGFR.
  \item[hsCRP] High sensitivity CRP.
  \item[Lab history] Structured list of results (test code, value, unit, date, lab name) when available.
\end{description}

\subsubsection{Health Conditions}
\paragraph{Non-communicable diseases}
\begin{itemize}
  \item NCD Model - Granularity supported. (For example, separate capture of type 1 diabetes, type 2 diabetes, diabetes onset age, and gestational diabetes history, hypertension, hyperlipidemia, ischemic heart disease, chronic kidney disease stage, liver disease, osteoporosis, thyroid conditions, cancer history with stage and treatment status, and disability descriptors).
\end{itemize}

\paragraph{Reproductive health}
\begin{itemize}
  \item Pregnancy status captured with structured fields such as current pregnancy flag, estimated due date, trimester, and breastfeeding status; contraception use and other reproductive metrics are modelled in more detail.
\end{itemize}

\paragraph{Mental health}
\begin{itemize}
  \item Mental health items can include structured assessments (for example, diagnosis flags with severity scores and last assessment dates for depression and anxiety) alongside traditional condition listings.
\end{itemize}

\paragraph{Immunological}
\begin{itemize}
  \item Autoimmune conditions, allergies (including reaction severity and whether confirmed by test), food intolerances.
\end{itemize}

\paragraph{Communicable \& nutritional}
\begin{itemize}
  \item History of communicable diseases and nutritional conditions can include richer status information (for example, hepatitis B vaccination/infection/recovery status, HIV status categories, and anemia with last hemoglobin value).
\end{itemize}

\subsubsection{Genetic Predispositions}
\begin{description}
  \item[Family history] Diabetes, cancer, heart disease, etc.
  \item[Genetic tests] BRCA, pharmacogenomic markers, other variants.
  \item[Pharmacogenomics profile] Relevant drug–gene interactions.
  \item[Notes] The updated model includes flags for specific predispositions (for example, family history of cardiovascular disease or known BRCA mutation) when available.
\end{description}

\subsection{Extrinsic Factors}

\subsubsection{Lifestyle Behaviors}
\begin{description}
  \item[diet] Preferences, restrictions, and fasting (including religious restrictions and fasting), food insecurity.
  \item[physical activity] Frequency, intensity, type.
  \item[substance use] Tobacco, alcohol, drugs.
  \item[sleep habits] Bedtime, wake time, disruptions.
\end{description}

\subsubsection{Psychosocial Context}
\begin{description}
  \item[social support] Caregivers, networks (May include household size, a list of close contacts, and caregiver entries with availability schedules).
  \item[digital health literacy] Low, moderate, high.
  \item[stigma/trust] Self-reported experiences.
  \item[relationships] Family structure, partner status.
  \item[care burden] Hours of unpaid caregiving.
\end{description}

\subsubsection{Environment}
\begin{description}
  \item[location] GPS or address (captured as structured address/GPS fields (country, state, city, pincode, lat/lon) when available).
  \item[housing] Quality, crowding, ownership.
  \item[air quality] PM2.5, AQI.
  \item[water \& sanitation] Access to clean water, toilets.
  \item[climate] Temperature, rainfall, seasonal hazards.
\end{description}

\subsubsection{Economic \& Health System}
\begin{description}
  \item[insurance coverage] Public, private, none.
  \item[government schemes] Enrollment status.
  \item[expenditures] Out-of-pocket costs.
  \item[transport access] To nearest facility.
  \item[nearest facility] Facility type, distance.
\end{description}

\subsubsection{Information Ecosystem}
\begin{description}
  \item[preferred sources] Radio, TV, community workers, internet (Recorded with relative weights and platform details.).
  \item[platforms] WhatsApp, Facebook, SMS.
  \item[verification] Fact-checking practices.
  \item[misinformation exposure] Frequency and type.
\end{description}

\subsubsection{Technology Access}
\begin{description}
  \item[devices] Smartphone, feature phone, tablet (Device inventory, connectivity type/reliability, and last-seen timestamps captured when available for richer device/telemetry support).
  \item[connectivity] 2G, 3G, 4G, WiFi.
  \item[telemedicine] Usage frequency.
  \item[digital accounts] IDs, logins.
\end{description}

\subsection{Care \& Interaction Factors}

\subsubsection{Roles \& Actors}
\begin{description}
  \item[actor registry] Mother, father, children, extended family, community health worker, midwife, nurse, physician, peers (includes structured attributes such as contact methods, default visibility, availability (including religious observances), delegation acceptance, and authority-level controls (e.g., initiation of escalation or visibility revocation)).
\end{description}

\subsubsection{Care Teams \& Care Plans}
\begin{description}
  \item[composition - care teams/care plans] (modeled explicitly (teams with member lists; care plans with goals, tasks, assigned actors, and review intervals).
  \item[goals] Health and wellness targets.
  \item[tasks] Daily, weekly, long-term.
\end{description}

\subsubsection{Responsibility \& State}
\begin{description}
  \item[state machine] Assignment, escalation, deferral rules (further captured with assigned actor, timestamps, deferral history, and escalation thresholds).
\end{description}

\subsubsection{Notification \& Preferences}
\begin{description}
  \item[channels] SMS, WhatsApp, IVR.
  \item[quiet hours] Time windows (Quiet hours can be a list of ranges, and notification preferences can include escalation contacts and max notifications per day (including religious observances)).
  \item[format] Text, audio.
  \item[languages] Preferred languages.
\end{description}

\subsubsection{Availability \& Presence}
\begin{description}
  \item[status] Active, idle, offline.
  \item[last activity] Timestamp.
  \item[confidence] High, medium, low (presence/confidence may be captured as a probabilistic or numeric score in addition to categorical status).
\end{description}

\subsubsection{Policy / Consent / Access Control}
\begin{description}
  \item[consent records] Given by whom, to whom, scope, duration.
  \item[access policies] Task-based, time-based modeled as structured records (including scope, parties, timestamps, restrictions, and revocation status) and policy rules with enforcement metadata.
  \item[deferral rules] Temporary delegation.
  \item[escalation] Thresholds and routing.
  \item[authority level] Escalation and visibility control.
\end{description}

\subsubsection{Provenance \& Audit}
\begin{description}
  \item[audit logs] Event-level tracking (Audit entries captured as structured events (actor, action type, target resource, reason, timestamp, evidence)).
  \item[provenance summaries] Who accessed what and why.
\end{description}

\subsubsection{Incident \& Harm Reporting}
\begin{description}
  \item[incident records] Safety events, reporting (Include severity, linked audit entries, mitigation actions, and governance review workflows).
  \item[governance reviews] Oversight outcomes.
\end{description}

\subsubsection{Renegotiation \& Governance Primitives}
\begin{description}
  \item[role review] Frequency and scope.
  \item[governance panels] Community or organizational (Panels can be modeled with membership and scope).
\end{description}

\subsubsection{Social \& Cultural Primitives}
\begin{description}
  \item[rituals] Context-specific care practices.
  \item[care practices] Cultural norms.
  \item[negotiation preferences] Approaches to shared decision making.
  \item[privacy granularity] Fine- or coarse-grained.
\end{description}

\subsubsection{Trust, Fairness \& Equity}
\begin{description}
  \item[fairness scores] Algorithmic fairness checks.
  \item[equity flags] Markers for disparities.
  \item[bias audit notes] Documentation of audit.
\end{description}

\subsubsection{Accessibility \& Assistive Needs}
\begin{description}
  \item[vision] Blindness, low vision, corrective aids.
  \item[hearing] Hearing loss, devices.
  \item[mobility] Wheelchair use, mobility aids.
  \item[supports] Assistive tech in use.
  \item[preferred a11y modes] Preferred accessibility modes (e.g., larger text, TTS, high contrast) can be captured and toggled.
\end{description}

\subsubsection{Data Governance \& Research}
\begin{description}
  \item[research consents] Granted/denied.
  \item[sharing preferences] Public, restricted, private.
  \item[retention] Duration of data retention.
\end{description}

\subsubsection{Device \& Telemetry}
\begin{description}
  \item[device status] Online/offline (Device inventory, battery/last-sync info, and detailed sensor metadata are captured in the updated model where available).
  \item[sensor metadata] Type, accuracy, calibration.
  \item[connectivity history] Logs.
\end{description}

\subsubsection{Emergency \& Safety}
\begin{description}
  \item[contacts] Emergency contact list.
  \item[directives] Advance care directives (Recorded with structured references (living will, power of attorney, document links, inclusive of religious practices)).
  \item[emergency plans] Household and care team protocols (include preferred hospital, transport method, and special instructions (inclusive of religious practices)).
\end{description}

\subsubsection{Derived \& Computed Fields}
\begin{description}
  \item[risk scores] Health risk indices.
  \item[engagement] Participation levels.
  \item[provenance traces] Derived data lineage.
\end{description}

\subsubsection{Administrative Data}
\begin{description}
  \item[record metadata] Versioning, timestamps.
  \item[clinical encounters] Visit records, provider notes.
\end{description}

\subsubsection{Administrative Data}
\begin{description}
  \item[record metadata] Versioning, timestamps, source (includes type of source of the record and confidence score).
  \item[medical and clinical encounters] All clinical information.
      \begin{description}
          \item [medication and allergies] Current list + history (including dose, frequency, dates prescribed, adherence history), and medication allergies (including allergen, reaction, severity, and test confirmation).
          \item [immunizations] vaccine, date, provider.
          \item [encounter history] facility, clinician, reason, outcome.
          \item [lab orders] tests, ordering provider, status.
          \item [clinical documents] type, date, link.
          \item [diagnostic and imaging] report, date, summary, link.
          \item [providers registry] name, organization, contact.
          \item [facilities registry] name, type, services, contact.
    \end{description}
\end{description}
\clearpage

\section{LLM Algorithms}
\label{appendix:algo}
The detailed algorithms for the verifier and summarizer are presented below.

\begin{algorithm}[ht!]
\caption{Algorithm for Health Information Verification Engine with User's Contextual Relevance}
\label{alg:context_builder}
\begin{algorithmic}[1]
  \Require user query $Q$, user id $u$, chat context $C$, profile store $\mathcal{P}$, care-graph $G$, retrieval size $k$
  \Ensure structured response context $CTX$ (snapshot, evidence IDs, prompts)

  \State $A \gets \text{ActorProfile}(u,G)$
  \State $Allowed \gets \text{ApplyACLs}(A,Q)$
  \State $Snapshot \gets \text{BuildSnapshot}(\mathcal{P},Allowed,C)$
  \Statex \hspace*{6em}$\triangleright$ \textit{privacy-limited}
  \State $E \gets \text{RetrieveTopK}(\text{Embed}(Snapshot),k)$
  \State $Sys \gets \text{SystemPersona}(A.role)$
  \State $User \gets \text{Concat}(Q,\text{ShortContext}(Snapshot),E)$
  \State $Draft \gets \text{LLM\_Generate}(\text{FormatPrompt}(Sys,User))$
  \State $Claims \gets \text{ExtractClaims}(Draft)$
  \ForAll{claim $c$ \textbf{in} $Claims$}
    \State $V \gets \text{LLM\_Verify}(c,\text{FindEvidence}(c,E,Snapshot))$
    \State $\text{AttachVerification}(Draft,c,V)$
  \EndFor
  \State $CTX \gets \text{SynthesizeContext}(Draft,E,Snapshot)$
  \State \Return $CTX$
\end{algorithmic}
\end{algorithm}

\begin{algorithm}[ht!]
\caption{AAlgorithm for Profile Summary Generation for Subject \& Care Circle Actors}
\label{alg:profile_summary}
\begin{algorithmic}[1]
  \Require canonical profile $\mathcal{P}$ for subject $s$, care circle $G_s$, templates $T$, time window $W$
  \Ensure per-actor summaries $\{S_a\}_{a\in G_s}$

  \State $Features \gets \text{EnrichFeatures}(\mathcal{P},W)$
  \State $S_s \gets \text{TemplateFill}(T[\text{subject}],Features)$
  \State $S_s.conf \gets \text{EstimateConfidence}(Features)$
  \State \text{StoreSnapshot}(s,$S_s$); \ \text{IndexEmbedding}(s,$S_s$)
  \ForAll{actor $a$ \textbf{in} $G_s$}
    \State $role \gets G_s.role(a)$
    \State $Perm \gets \text{ResolvePermissions}(a,\mathcal{P},G_s)$
    \State $Fields \gets \text{FilterFields}(Features,Perm)$
    \State $brief \gets \text{TemplateFill}(T[role],Fields)$
    \State $brief.actionable \gets \text{ExtractTasks}(brief)$
    \State \text{AnnotateProvenance}(brief)
    \State \text{StoreSnapshot}(a,brief); \ \text{IndexEmbedding}(a,brief)
    \State \text{SendNotificationIfNeeded}(a,brief)
  \EndFor
  \State \Return $\{S_a\}_{a\in G_s}$
\end{algorithmic}
\end{algorithm}

%% file: template.bbl

\begin{thebibliography}{69}


\ifx \showCODEN    \undefined \def \showCODEN     #1{\unskip}     \fi
\ifx \showDOI      \undefined \def \showDOI       #1{#1}\fi
\ifx \showISBNx    \undefined \def \showISBNx     #1{\unskip}     \fi
\ifx \showISBNxiii \undefined \def \showISBNxiii  #1{\unskip}     \fi
\ifx \showISSN     \undefined \def \showISSN      #1{\unskip}     \fi
\ifx \showLCCN     \undefined \def \showLCCN      #1{\unskip}     \fi
\ifx \shownote     \undefined \def \shownote      #1{#1}          \fi
\ifx \showarticletitle \undefined \def \showarticletitle #1{#1}   \fi
\ifx \showURL      \undefined \def \showURL       {\relax}        \fi
\providecommand\bibfield[2]{#2}
\providecommand\bibinfo[2]{#2}
\providecommand\natexlab[1]{#1}
\providecommand\showeprint[2][]{arXiv:#2}

\bibitem[Abdelaziz et~al\mbox{.}(2024)]%
        {Abdelaziz_Garfield_Neves_Lloyd_Norton_van_Dael_Wheeler_McLeod_Franklin_2024}
\bibfield{author}{\bibinfo{person}{Shahd Abdelaziz}, \bibinfo{person}{Sara Garfield}, \bibinfo{person}{Ana~Luisa Neves}, \bibinfo{person}{Jill Lloyd}, \bibinfo{person}{John Norton}, \bibinfo{person}{Jackie van Dael}, \bibinfo{person}{Carly Wheeler}, \bibinfo{person}{Monsey McLeod}, {and} \bibinfo{person}{Bryony~Dean Franklin}.} \bibinfo{year}{2024}\natexlab{}.
\newblock \showarticletitle{What are the unintended patient safety consequences of healthcare technologies? A qualitative study among patients, carers and healthcare providers}.
\newblock \bibinfo{journal}{\emph{BMJ Open}} \bibinfo{volume}{14}, \bibinfo{number}{11} (\bibinfo{date}{Nov} \bibinfo{year}{2024}), \bibinfo{pages}{1–10}.
\newblock
\urldef\tempurl%
\url{https://doi.org/10.1136/bmjopen-2024-089026}
\showDOI{\tempurl}


\bibitem[Ackerman(2000)]%
        {10.1207/S15327051HCI1523_5}
\bibfield{author}{\bibinfo{person}{Mark~S. Ackerman}.} \bibinfo{year}{2000}\natexlab{}.
\newblock \showarticletitle{The intellectual challenge of CSCW: the gap between social requirements and technical feasibility}.
\newblock \bibinfo{journal}{\emph{Hum.-Comput. Interact.}} \bibinfo{volume}{15}, \bibinfo{number}{2} (\bibinfo{date}{Sept.} \bibinfo{year}{2000}), \bibinfo{pages}{179–203}.
\newblock
\showISSN{0737-0024}
\urldef\tempurl%
\url{https://doi.org/10.1207/S15327051HCI1523_5}
\showDOI{\tempurl}


\bibitem[Amershi et~al\mbox{.}(2019)]%
        {10.1145/3290605.3300233}
\bibfield{author}{\bibinfo{person}{Saleema Amershi}, \bibinfo{person}{Dan Weld}, \bibinfo{person}{Mihaela Vorvoreanu}, \bibinfo{person}{Adam Fourney}, \bibinfo{person}{Besmira Nushi}, \bibinfo{person}{Penny Collisson}, \bibinfo{person}{Jina Suh}, \bibinfo{person}{Shamsi Iqbal}, \bibinfo{person}{Paul~N. Bennett}, \bibinfo{person}{Kori Inkpen}, \bibinfo{person}{Jaime Teevan}, \bibinfo{person}{Ruth Kikin-Gil}, {and} \bibinfo{person}{Eric Horvitz}.} \bibinfo{year}{2019}\natexlab{}.
\newblock \showarticletitle{Guidelines for Human-AI Interaction}. In \bibinfo{booktitle}{\emph{Proceedings of the 2019 CHI Conference on Human Factors in Computing Systems}} (Glasgow, Scotland Uk) \emph{(\bibinfo{series}{CHI '19})}. \bibinfo{publisher}{Association for Computing Machinery}, \bibinfo{address}{New York, NY, USA}, \bibinfo{pages}{1–13}.
\newblock
\showISBNx{9781450359702}
\urldef\tempurl%
\url{https://doi.org/10.1145/3290605.3300233}
\showDOI{\tempurl}


\bibitem[Anvari(2025)]%
        {10.1145/3715668.3735626}
\bibfield{author}{\bibinfo{person}{Soraya~S. Anvari}.} \bibinfo{year}{2025}\natexlab{}.
\newblock \bibinfo{booktitle}{\emph{Designing Interactive Artifacts for Mental Health Education: A Game-Based Approach using AI as In-game Characters}}.
\newblock \bibinfo{publisher}{Association for Computing Machinery}, \bibinfo{address}{New York, NY, USA}, \bibinfo{pages}{82–85}.
\newblock
\showISBNx{9798400714863}
\urldef\tempurl%
\url{https://doi.org/10.1145/3715668.3735626}
\showURL{%
\tempurl}


\bibitem[Badillo-Urquiola et~al\mbox{.}(2018)]%
        {10.1145/3272973.3273012}
\bibfield{author}{\bibinfo{person}{Karla Badillo-Urquiola}, \bibinfo{person}{Yaxing Yao}, \bibinfo{person}{Oshrat Ayalon}, \bibinfo{person}{Bart Knijnenurg}, \bibinfo{person}{Xinru Page}, \bibinfo{person}{Eran Toch}, \bibinfo{person}{Yang Wang}, {and} \bibinfo{person}{Pamela~J. Wisniewski}.} \bibinfo{year}{2018}\natexlab{}.
\newblock \showarticletitle{Privacy in Context: Critically Engaging with Theory to Guide Privacy Research and Design}. In \bibinfo{booktitle}{\emph{Companion of the 2018 ACM Conference on Computer Supported Cooperative Work and Social Computing}} (Jersey City, NJ, USA) \emph{(\bibinfo{series}{CSCW '18 Companion})}. \bibinfo{publisher}{Association for Computing Machinery}, \bibinfo{address}{New York, NY, USA}, \bibinfo{pages}{425–431}.
\newblock
\showISBNx{9781450360180}
\urldef\tempurl%
\url{https://doi.org/10.1145/3272973.3273012}
\showDOI{\tempurl}


\bibitem[Bai et~al\mbox{.}(2021)]%
        {10.1145/3460421.3480429}
\bibfield{author}{\bibinfo{person}{Yang Bai}, \bibinfo{person}{Yu Guan}, \bibinfo{person}{Jian~Qing Shi}, {and} \bibinfo{person}{Wan‑Fai Ng}.} \bibinfo{year}{2021}\natexlab{}.
\newblock \showarticletitle{Towards Automated Fatigue Assessment using Wearable Sensing and Mixed-Effects Models}. In \bibinfo{booktitle}{\emph{Proceedings of the 2021 ACM International Symposium on Wearable Computers}} (Virtual, USA) \emph{(\bibinfo{series}{ISWC '21})}. \bibinfo{publisher}{Association for Computing Machinery}, \bibinfo{address}{New York, NY, USA}, \bibinfo{pages}{129–131}.
\newblock
\showISBNx{9781450384629}
\urldef\tempurl%
\url{https://doi.org/10.1145/3460421.3480429}
\showDOI{\tempurl}


\bibitem[Balaji et~al\mbox{.}(2019)]%
        {10.1145/3307334.3328583}
\bibfield{author}{\bibinfo{person}{Ananta~Narayanan Balaji}, \bibinfo{person}{Chen Yuan}, \bibinfo{person}{Bo Wang}, \bibinfo{person}{Li-Shiuan Peh}, {and} \bibinfo{person}{Huilin Shao}.} \bibinfo{year}{2019}\natexlab{}.
\newblock \showarticletitle{pH Watch - Leveraging Pulse Oximeters in Existing Wearables for Reusable, Real-time Monitoring of pH in Sweat (demo)}. In \bibinfo{booktitle}{\emph{Proceedings of the 17th Annual International Conference on Mobile Systems, Applications, and Services}} (Seoul, Republic of Korea) \emph{(\bibinfo{series}{MobiSys '19})}. \bibinfo{publisher}{Association for Computing Machinery}, \bibinfo{address}{New York, NY, USA}, \bibinfo{pages}{687–688}.
\newblock
\showISBNx{9781450366618}
\urldef\tempurl%
\url{https://doi.org/10.1145/3307334.3328583}
\showDOI{\tempurl}


\bibitem[Bogdanova(2024)]%
        {10.1145/3656156.3665127}
\bibfield{author}{\bibinfo{person}{Karin Bogdanova}.} \bibinfo{year}{2024}\natexlab{}.
\newblock \showarticletitle{Aesthetics of algorithmic care: Designing alternative human-AI collaboration practices for digital phenotyping}. In \bibinfo{booktitle}{\emph{Companion Publication of the 2024 ACM Designing Interactive Systems Conference}} (IT University of Copenhagen, Denmark) \emph{(\bibinfo{series}{DIS '24 Companion})}. \bibinfo{publisher}{Association for Computing Machinery}, \bibinfo{address}{New York, NY, USA}, \bibinfo{pages}{59–61}.
\newblock
\showISBNx{9798400706325}
\urldef\tempurl%
\url{https://doi.org/10.1145/3656156.3665127}
\showDOI{\tempurl}


\bibitem[Cajamarca et~al\mbox{.}(2023)]%
        {10.1145/3563657.3596109}
\bibfield{author}{\bibinfo{person}{Gabriela Cajamarca}, \bibinfo{person}{Valeria Herskovic}, \bibinfo{person}{Stephannie Dondighual}, \bibinfo{person}{Carolina Fuentes}, {and} \bibinfo{person}{Nervo Verdezoto}.} \bibinfo{year}{2023}\natexlab{}.
\newblock \showarticletitle{Understanding how to Design Health Data Visualizations for Chilean Older Adults on Mobile Devices}. In \bibinfo{booktitle}{\emph{Proceedings of the 2023 ACM Designing Interactive Systems Conference}} (Pittsburgh, PA, USA) \emph{(\bibinfo{series}{DIS '23})}. \bibinfo{publisher}{Association for Computing Machinery}, \bibinfo{address}{New York, NY, USA}, \bibinfo{pages}{1309–1324}.
\newblock
\showISBNx{9781450398930}
\urldef\tempurl%
\url{https://doi.org/10.1145/3563657.3596109}
\showDOI{\tempurl}


\bibitem[Chen et~al\mbox{.}(2024)]%
        {10.1145/3636534.3649366}
\bibfield{author}{\bibinfo{person}{Tao Chen}, \bibinfo{person}{Yongjie Yang}, \bibinfo{person}{Xiaoran Fan}, \bibinfo{person}{Xiuzhen Guo}, \bibinfo{person}{Jie Xiong}, {and} \bibinfo{person}{Longfei Shangguan}.} \bibinfo{year}{2024}\natexlab{}.
\newblock \showarticletitle{Exploring the Feasibility of Remote Cardiac Auscultation Using Earphones}. In \bibinfo{booktitle}{\emph{Proceedings of the 30th Annual International Conference on Mobile Computing and Networking}} (Washington D.C., DC, USA) \emph{(\bibinfo{series}{ACM MobiCom '24})}. \bibinfo{publisher}{Association for Computing Machinery}, \bibinfo{address}{New York, NY, USA}, \bibinfo{pages}{357–372}.
\newblock
\showISBNx{9798400704895}
\urldef\tempurl%
\url{https://doi.org/10.1145/3636534.3649366}
\showDOI{\tempurl}


\bibitem[Chen et~al\mbox{.}(2020)]%
        {10.1145/3384419.3430442}
\bibfield{author}{\bibinfo{person}{Xingyu Chen}, \bibinfo{person}{Chenhan Xu}, \bibinfo{person}{Baicheng Chen}, \bibinfo{person}{Zhengxiong Li}, {and} \bibinfo{person}{Wenyao Xu}.} \bibinfo{year}{2020}\natexlab{}.
\newblock \showarticletitle{In-ear thermometer: wearable real-time core body temperature monitoring: poster abstract}. In \bibinfo{booktitle}{\emph{Proceedings of the 18th Conference on Embedded Networked Sensor Systems}} (Virtual Event, Japan) \emph{(\bibinfo{series}{SenSys '20})}. \bibinfo{publisher}{Association for Computing Machinery}, \bibinfo{address}{New York, NY, USA}, \bibinfo{pages}{687–688}.
\newblock
\showISBNx{9781450375900}
\urldef\tempurl%
\url{https://doi.org/10.1145/3384419.3430442}
\showDOI{\tempurl}


\bibitem[Chen et~al\mbox{.}(2013)]%
        {10.1145/2441776.2441789}
\bibfield{author}{\bibinfo{person}{Yunan Chen}, \bibinfo{person}{Victor Ngo}, {and} \bibinfo{person}{Sun~Young Park}.} \bibinfo{year}{2013}\natexlab{}.
\newblock \showarticletitle{Caring for caregivers: designing for integrality}. In \bibinfo{booktitle}{\emph{Proceedings of the 2013 Conference on Computer Supported Cooperative Work}} (San Antonio, Texas, USA) \emph{(\bibinfo{series}{CSCW '13})}. \bibinfo{publisher}{Association for Computing Machinery}, \bibinfo{address}{New York, NY, USA}, \bibinfo{pages}{91–102}.
\newblock
\showISBNx{9781450313315}
\urldef\tempurl%
\url{https://doi.org/10.1145/2441776.2441789}
\showDOI{\tempurl}


\bibitem[Choube et~al\mbox{.}(2025)]%
        {10.1145/3749474}
\bibfield{author}{\bibinfo{person}{Akshat Choube}, \bibinfo{person}{Ha Le}, \bibinfo{person}{Jiachen Li}, \bibinfo{person}{Kaixin Ji}, \bibinfo{person}{Vedant~Das Swain}, {and} \bibinfo{person}{Varun Mishra}.} \bibinfo{year}{2025}\natexlab{}.
\newblock \showarticletitle{GLOSS: Group of LLMs for Open-ended Sensemaking of Passive Sensing Data for Health and Wellbeing}.
\newblock \bibinfo{journal}{\emph{Proc. ACM Interact. Mob. Wearable Ubiquitous Technol.}} \bibinfo{volume}{9}, \bibinfo{number}{3}, Article \bibinfo{articleno}{76} (\bibinfo{date}{Sept.} \bibinfo{year}{2025}), \bibinfo{numpages}{32}~pages.
\newblock
\urldef\tempurl%
\url{https://doi.org/10.1145/3749474}
\showDOI{\tempurl}


\bibitem[Cornelius et~al\mbox{.}(2014)]%
        {10.1145/2594368.2594369}
\bibfield{author}{\bibinfo{person}{Cory Cornelius}, \bibinfo{person}{Ronald Peterson}, \bibinfo{person}{Joseph Skinner}, \bibinfo{person}{Ryan Halter}, {and} \bibinfo{person}{David Kotz}.} \bibinfo{year}{2014}\natexlab{}.
\newblock \showarticletitle{A wearable system that knows who wears it}. In \bibinfo{booktitle}{\emph{Proceedings of the 12th Annual International Conference on Mobile Systems, Applications, and Services}} (Bretton Woods, New Hampshire, USA) \emph{(\bibinfo{series}{MobiSys '14})}. \bibinfo{publisher}{Association for Computing Machinery}, \bibinfo{address}{New York, NY, USA}, \bibinfo{pages}{55–67}.
\newblock
\showISBNx{9781450327930}
\urldef\tempurl%
\url{https://doi.org/10.1145/2594368.2594369}
\showDOI{\tempurl}


\bibitem[Curtis et~al\mbox{.}(2019)]%
        {10.1145/3300061.3345433}
\bibfield{author}{\bibinfo{person}{Aidan Curtis}, \bibinfo{person}{Amruta Pai}, \bibinfo{person}{Jian Cao}, \bibinfo{person}{Nidal Moukaddam}, {and} \bibinfo{person}{Ashutosh Sabharwal}.} \bibinfo{year}{2019}\natexlab{}.
\newblock \showarticletitle{HealthSense: Software-defined Mobile-based Clinical Trials}. In \bibinfo{booktitle}{\emph{The 25th Annual International Conference on Mobile Computing and Networking}} (Los Cabos, Mexico) \emph{(\bibinfo{series}{MobiCom '19})}. \bibinfo{publisher}{Association for Computing Machinery}, \bibinfo{address}{New York, NY, USA}, Article \bibinfo{articleno}{32}, \bibinfo{numpages}{15}~pages.
\newblock
\showISBNx{9781450361699}
\urldef\tempurl%
\url{https://doi.org/10.1145/3300061.3345433}
\showDOI{\tempurl}


\bibitem[D'Adamo et~al\mbox{.}(2024)]%
        {10.1145/3613904.3642651}
\bibfield{author}{\bibinfo{person}{Amar D'Adamo}, \bibinfo{person}{Marte Roel~Lesur}, \bibinfo{person}{Laia Turmo~Vidal}, \bibinfo{person}{Mohammad~Mahdi Dehshibi}, \bibinfo{person}{Daniel De~La~Prida}, \bibinfo{person}{Joaqu\'{\i}n~R. Diaz-Dur\'{a}n}, \bibinfo{person}{Luis~Antonio Azpicueta-Ruiz}, \bibinfo{person}{Aleksander V\"{a}ljam\"{a}e}, {and} \bibinfo{person}{Ana Tajadura-Jim\'{e}nez}.} \bibinfo{year}{2024}\natexlab{}.
\newblock \showarticletitle{SoniWeight Shoes: Investigating Effects and Personalization of a Wearable Sound Device for Altering Body Perception and Behavior}. In \bibinfo{booktitle}{\emph{Proceedings of the 2024 CHI Conference on Human Factors in Computing Systems}} (Honolulu, HI, USA) \emph{(\bibinfo{series}{CHI '24})}. \bibinfo{publisher}{Association for Computing Machinery}, \bibinfo{address}{New York, NY, USA}, Article \bibinfo{articleno}{93}, \bibinfo{numpages}{20}~pages.
\newblock
\showISBNx{9798400703300}
\urldef\tempurl%
\url{https://doi.org/10.1145/3613904.3642651}
\showDOI{\tempurl}


\bibitem[Debackere et~al\mbox{.}(2025)]%
        {10.1145/3706598.3713224}
\bibfield{author}{\bibinfo{person}{Florian Debackere}, \bibinfo{person}{C\'{e}line Clavel}, \bibinfo{person}{Alexandra R\"{o}ren}, \bibinfo{person}{Fran\c{c}ois Rannou}, \bibinfo{person}{Christelle Nguyen}, \bibinfo{person}{Viet-thi Tran}, \bibinfo{person}{Yosra Messai}, {and} \bibinfo{person}{Jean-Claude Martin}.} \bibinfo{year}{2025}\natexlab{}.
\newblock \showarticletitle{Evaluation of a Tailored Mobile Application for Self-Management of Low Back Pain: Towards a Metamodel for Designing Behavior Change Technologies}. In \bibinfo{booktitle}{\emph{Proceedings of the 2025 CHI Conference on Human Factors in Computing Systems}} (Yokohama, Japan) \emph{(\bibinfo{series}{CHI '25})}. \bibinfo{publisher}{Association for Computing Machinery}, \bibinfo{address}{New York, NY, USA}, Article \bibinfo{articleno}{516}, \bibinfo{numpages}{17}~pages.
\newblock
\showISBNx{9798400713941}
\urldef\tempurl%
\url{https://doi.org/10.1145/3706598.3713224}
\showDOI{\tempurl}


\bibitem[Ehsan and Riedl(2020)]%
        {10.1007/978-3-030-60117-1_33}
\bibfield{author}{\bibinfo{person}{Upol Ehsan} {and} \bibinfo{person}{Mark~O. Riedl}.} \bibinfo{year}{2020}\natexlab{}.
\newblock \showarticletitle{Human-Centered Explainable AI: Towards a Reflective Sociotechnical Approach}. In \bibinfo{booktitle}{\emph{HCI International 2020 - Late Breaking Papers: Multimodality and Intelligence: 22nd HCI International Conference, HCII 2020, Copenhagen, Denmark, July 19–24, 2020, Proceedings}} (Copenhagen, Denmark). \bibinfo{publisher}{Springer-Verlag}, \bibinfo{address}{Berlin, Heidelberg}, \bibinfo{pages}{449–466}.
\newblock
\showISBNx{978-3-030-60116-4}
\urldef\tempurl%
\url{https://doi.org/10.1007/978-3-030-60117-1_33}
\showDOI{\tempurl}


\bibitem[Ekhtiar et~al\mbox{.}(2025)]%
        {10.1145/3715336.3735813}
\bibfield{author}{\bibinfo{person}{Tina Ekhtiar}, \bibinfo{person}{Arma\u{g}an Karahano\u{g}lu}, \bibinfo{person}{Ruben Gouveia}, {and} \bibinfo{person}{Geke~D.S. Ludden}.} \bibinfo{year}{2025}\natexlab{}.
\newblock \showarticletitle{Changing Health Goals with Personal Informatics}. In \bibinfo{booktitle}{\emph{Proceedings of the 2025 ACM Designing Interactive Systems Conference}} \emph{(\bibinfo{series}{DIS '25})}. \bibinfo{publisher}{Association for Computing Machinery}, \bibinfo{address}{New York, NY, USA}, \bibinfo{pages}{2867–2883}.
\newblock
\showISBNx{9798400714856}
\urldef\tempurl%
\url{https://doi.org/10.1145/3715336.3735813}
\showDOI{\tempurl}


\bibitem[Ertin et~al\mbox{.}(2011)]%
        {10.1145/2070942.2071027}
\bibfield{author}{\bibinfo{person}{Emre Ertin}, \bibinfo{person}{Andrew Raij}, \bibinfo{person}{Nathan Stohs}, \bibinfo{person}{Mustafa al'Absi}, \bibinfo{person}{Santosh Kumar}, {and} \bibinfo{person}{Somnath Mitra}.} \bibinfo{year}{2011}\natexlab{}.
\newblock \showarticletitle{Demo: An unobtrusively wearable sensor suite for inferring the onset, causality, and consequences of stress in the field}. In \bibinfo{booktitle}{\emph{Proceedings of the 9th ACM Conference on Embedded Networked Sensor Systems}} (Seattle, Washington) \emph{(\bibinfo{series}{SenSys '11})}. \bibinfo{publisher}{Association for Computing Machinery}, \bibinfo{address}{New York, NY, USA}, \bibinfo{pages}{437–438}.
\newblock
\showISBNx{9781450307185}
\urldef\tempurl%
\url{https://doi.org/10.1145/2070942.2071027}
\showDOI{\tempurl}


\bibitem[Fan et~al\mbox{.}(2023)]%
        {10.1145/3570361.3613281}
\bibfield{author}{\bibinfo{person}{Xiaoran Fan}, \bibinfo{person}{David Pearl}, \bibinfo{person}{Richard Howard}, \bibinfo{person}{Longfei Shangguan}, {and} \bibinfo{person}{Trausti Thormundsson}.} \bibinfo{year}{2023}\natexlab{}.
\newblock \showarticletitle{APG: Audioplethysmography for Cardiac Monitoring in Hearables}. In \bibinfo{booktitle}{\emph{Proceedings of the 29th Annual International Conference on Mobile Computing and Networking}} (Madrid, Spain) \emph{(\bibinfo{series}{ACM MobiCom '23})}. \bibinfo{publisher}{Association for Computing Machinery}, \bibinfo{address}{New York, NY, USA}, Article \bibinfo{articleno}{67}, \bibinfo{numpages}{15}~pages.
\newblock
\showISBNx{9781450399906}
\urldef\tempurl%
\url{https://doi.org/10.1145/3570361.3613281}
\showDOI{\tempurl}


\bibitem[F\'{e}lix et~al\mbox{.}(2024)]%
        {10.1145/3643834.3661504}
\bibfield{author}{\bibinfo{person}{Beatriz F\'{e}lix}, \bibinfo{person}{Cristiana Braga}, \bibinfo{person}{Xolani Ntinga}, \bibinfo{person}{Sarina~C Till}, \bibinfo{person}{Leina Meoli}, \bibinfo{person}{Alastair Van~Heerden}, \bibinfo{person}{Ricardo Melo}, \bibinfo{person}{Nervo Verdezoto}, \bibinfo{person}{Melissa Densmore}, {and} \bibinfo{person}{Francisco Nunes}.} \bibinfo{year}{2024}\natexlab{}.
\newblock \showarticletitle{Understanding How Parents Deal With the Health Advice They Receive: A Qualitative Study and Implications for the Design of Message-based Health Dissemination Systems for Child Health}. In \bibinfo{booktitle}{\emph{Proceedings of the 2024 ACM Designing Interactive Systems Conference}} (Copenhagen, Denmark) \emph{(\bibinfo{series}{DIS '24})}. \bibinfo{publisher}{Association for Computing Machinery}, \bibinfo{address}{New York, NY, USA}, \bibinfo{pages}{1319–1335}.
\newblock
\showISBNx{9798400705830}
\urldef\tempurl%
\url{https://doi.org/10.1145/3643834.3661504}
\showDOI{\tempurl}


\bibitem[Ganesan(2015)]%
        {10.1145/2801694.2801710}
\bibfield{author}{\bibinfo{person}{Deepak Ganesan}.} \bibinfo{year}{2015}\natexlab{}.
\newblock \showarticletitle{Towards Ultra-low Power Wearable Health Sensing with Sparse Sampling and Asymmetric Communication}. In \bibinfo{booktitle}{\emph{Proceedings of the 2015 Workshop on Wireless of the Students, by the Students, \& for the Students}} (Paris, France) \emph{(\bibinfo{series}{S3 '15})}. \bibinfo{publisher}{Association for Computing Machinery}, \bibinfo{address}{New York, NY, USA}, \bibinfo{pages}{34}.
\newblock
\showISBNx{9781450337014}
\urldef\tempurl%
\url{https://doi.org/10.1145/2801694.2801710}
\showDOI{\tempurl}


\bibitem[Helms and Fernaeus(2021)]%
        {10.1145/3461778.3462025}
\bibfield{author}{\bibinfo{person}{Karey Helms} {and} \bibinfo{person}{Ylva Fernaeus}.} \bibinfo{year}{2021}\natexlab{}.
\newblock \showarticletitle{Troubling Care: Four Orientations for Wickedness in Design}. In \bibinfo{booktitle}{\emph{Proceedings of the 2021 ACM Designing Interactive Systems Conference}} (Virtual Event, USA) \emph{(\bibinfo{series}{DIS '21})}. \bibinfo{publisher}{Association for Computing Machinery}, \bibinfo{address}{New York, NY, USA}, \bibinfo{pages}{789–801}.
\newblock
\showISBNx{9781450384766}
\urldef\tempurl%
\url{https://doi.org/10.1145/3461778.3462025}
\showDOI{\tempurl}


\bibitem[Hollan et~al\mbox{.}(2000)]%
        {10.1145/353485.353487}
\bibfield{author}{\bibinfo{person}{James Hollan}, \bibinfo{person}{Edwin Hutchins}, {and} \bibinfo{person}{David Kirsh}.} \bibinfo{year}{2000}\natexlab{}.
\newblock \showarticletitle{Distributed cognition: toward a new foundation for human-computer interaction research}.
\newblock \bibinfo{journal}{\emph{ACM Trans. Comput.-Hum. Interact.}} \bibinfo{volume}{7}, \bibinfo{number}{2} (\bibinfo{date}{June} \bibinfo{year}{2000}), \bibinfo{pages}{174–196}.
\newblock
\showISSN{1073-0516}
\urldef\tempurl%
\url{https://doi.org/10.1145/353485.353487}
\showDOI{\tempurl}


\bibitem[Hsieh et~al\mbox{.}(2014)]%
        {10.1145/2559206.2560474}
\bibfield{author}{\bibinfo{person}{Gary Hsieh}, \bibinfo{person}{Sean~A. Munson}, \bibinfo{person}{Maurits~C. Kaptein}, \bibinfo{person}{Harri Oinas-Kukkonen}, {and} \bibinfo{person}{Oded Nov}.} \bibinfo{year}{2014}\natexlab{}.
\newblock \showarticletitle{Personalizing behavior change technologies}. In \bibinfo{booktitle}{\emph{CHI '14 Extended Abstracts on Human Factors in Computing Systems}} (Toronto, Ontario, Canada) \emph{(\bibinfo{series}{CHI EA '14})}. \bibinfo{publisher}{Association for Computing Machinery}, \bibinfo{address}{New York, NY, USA}, \bibinfo{pages}{107–110}.
\newblock
\showISBNx{9781450324748}
\urldef\tempurl%
\url{https://doi.org/10.1145/2559206.2560474}
\showDOI{\tempurl}


\bibitem[Hsu et~al\mbox{.}(2025)]%
        {10.1145/3715336.3735678}
\bibfield{author}{\bibinfo{person}{Long-Jing Hsu}, \bibinfo{person}{Janice Bays}, \bibinfo{person}{Manasi Swaminathan}, \bibinfo{person}{Weslie Khoo}, \bibinfo{person}{Hiroki Sato}, \bibinfo{person}{Kyrie~Jig Amon}, \bibinfo{person}{Sathvika Dobbala}, \bibinfo{person}{Min~Min Thant}, \bibinfo{person}{Alex Foster}, \bibinfo{person}{Kate Tsui}, \bibinfo{person}{Philip~B. Stafford}, \bibinfo{person}{David Crandall}, {and} \bibinfo{person}{Selma Sabanovic}.} \bibinfo{year}{2025}\natexlab{}.
\newblock \showarticletitle{Research as Care: A Reflection on Incorporating the Ethics of Care in Design Research with People Living with Dementia}. In \bibinfo{booktitle}{\emph{Proceedings of the 2025 ACM Designing Interactive Systems Conference}} \emph{(\bibinfo{series}{DIS '25})}. \bibinfo{publisher}{Association for Computing Machinery}, \bibinfo{address}{New York, NY, USA}, \bibinfo{pages}{3013–3027}.
\newblock
\showISBNx{9798400714856}
\urldef\tempurl%
\url{https://doi.org/10.1145/3715336.3735678}
\showDOI{\tempurl}


\bibitem[Jang et~al\mbox{.}(2025)]%
        {10.1145/3706598.3714208}
\bibfield{author}{\bibinfo{person}{Sueun Jang}, \bibinfo{person}{Youngseok Seo}, \bibinfo{person}{Woohyeok Choi}, {and} \bibinfo{person}{Uichin Lee}.} \bibinfo{year}{2025}\natexlab{}.
\newblock \showarticletitle{Like Adding a Small Weight to a Scale About to Tip: Personalizing Micro-Financial Incentives for Digital Wellbeing}. In \bibinfo{booktitle}{\emph{Proceedings of the 2025 CHI Conference on Human Factors in Computing Systems}} (Yokohama, Japan) \emph{(\bibinfo{series}{CHI '25})}. \bibinfo{publisher}{Association for Computing Machinery}, \bibinfo{address}{New York, NY, USA}, Article \bibinfo{articleno}{1188}, \bibinfo{numpages}{19}~pages.
\newblock
\showISBNx{9798400713941}
\urldef\tempurl%
\url{https://doi.org/10.1145/3706598.3714208}
\showDOI{\tempurl}


\bibitem[Jansen et~al\mbox{.}(2020)]%
        {10.1145/3357236.3395469}
\bibfield{author}{\bibinfo{person}{Jos-Marien Jansen}, \bibinfo{person}{Karin Niemantsverdriet}, \bibinfo{person}{Anne~Wil Burghoorn}, \bibinfo{person}{Peter Lovei}, \bibinfo{person}{Ineke Neutelings}, \bibinfo{person}{Eva Deckers}, {and} \bibinfo{person}{Simon Nienhuijs}.} \bibinfo{year}{2020}\natexlab{}.
\newblock \showarticletitle{Design for Co-responsibility: Connecting Patients, Partners, and Professionals in Bariatric Lifestyle Changes}. In \bibinfo{booktitle}{\emph{Proceedings of the 2020 ACM Designing Interactive Systems Conference}} (Eindhoven, Netherlands) \emph{(\bibinfo{series}{DIS '20})}. \bibinfo{publisher}{Association for Computing Machinery}, \bibinfo{address}{New York, NY, USA}, \bibinfo{pages}{1537–1549}.
\newblock
\showISBNx{9781450369749}
\urldef\tempurl%
\url{https://doi.org/10.1145/3357236.3395469}
\showDOI{\tempurl}


\bibitem[Kawakami et~al\mbox{.}(2023)]%
        {10.1145/3610207}
\bibfield{author}{\bibinfo{person}{Anna Kawakami}, \bibinfo{person}{Shreya Chowdhary}, \bibinfo{person}{Shamsi~T. Iqbal}, \bibinfo{person}{Q.~Vera Liao}, \bibinfo{person}{Alexandra Olteanu}, \bibinfo{person}{Jina Suh}, {and} \bibinfo{person}{Koustuv Saha}.} \bibinfo{year}{2023}\natexlab{}.
\newblock \showarticletitle{Sensing Wellbeing in the Workplace, Why and For Whom? Envisioning Impacts with Organizational Stakeholders}.
\newblock \bibinfo{journal}{\emph{Proc. ACM Hum.-Comput. Interact.}} \bibinfo{volume}{7}, \bibinfo{number}{CSCW2}, Article \bibinfo{articleno}{358} (\bibinfo{date}{Oct.} \bibinfo{year}{2023}), \bibinfo{numpages}{33}~pages.
\newblock
\urldef\tempurl%
\url{https://doi.org/10.1145/3610207}
\showDOI{\tempurl}


\bibitem[Kaziunas et~al\mbox{.}(2017)]%
        {10.1145/2998181.2998303}
\bibfield{author}{\bibinfo{person}{Elizabeth Kaziunas}, \bibinfo{person}{Mark~S. Ackerman}, \bibinfo{person}{Silvia Lindtner}, {and} \bibinfo{person}{Joyce~M. Lee}.} \bibinfo{year}{2017}\natexlab{}.
\newblock \showarticletitle{Caring through Data: Attending to the Social and Emotional Experiences of Health Datafication}. In \bibinfo{booktitle}{\emph{Proceedings of the 2017 ACM Conference on Computer Supported Cooperative Work and Social Computing}} (Portland, Oregon, USA) \emph{(\bibinfo{series}{CSCW '17})}. \bibinfo{publisher}{Association for Computing Machinery}, \bibinfo{address}{New York, NY, USA}, \bibinfo{pages}{2260–2272}.
\newblock
\showISBNx{9781450343350}
\urldef\tempurl%
\url{https://doi.org/10.1145/2998181.2998303}
\showDOI{\tempurl}


\bibitem[Kovacs et~al\mbox{.}(2021)]%
        {10.1145/3411764.3445695}
\bibfield{author}{\bibinfo{person}{Geza Kovacs}, \bibinfo{person}{Zhengxuan Wu}, {and} \bibinfo{person}{Michael~S. Bernstein}.} \bibinfo{year}{2021}\natexlab{}.
\newblock \showarticletitle{Not Now, Ask Later: Users Weaken Their Behavior Change Regimen Over Time, But Expect To Re-Strengthen It Imminently}. In \bibinfo{booktitle}{\emph{Proceedings of the 2021 CHI Conference on Human Factors in Computing Systems}} (Yokohama, Japan) \emph{(\bibinfo{series}{CHI '21})}. \bibinfo{publisher}{Association for Computing Machinery}, \bibinfo{address}{New York, NY, USA}, Article \bibinfo{articleno}{229}, \bibinfo{numpages}{14}~pages.
\newblock
\showISBNx{9781450380966}
\urldef\tempurl%
\url{https://doi.org/10.1145/3411764.3445695}
\showDOI{\tempurl}


\bibitem[Krishnamoorthy~Srinivasan et~al\mbox{.}(2025)]%
        {10.1145/3706599.3720208}
\bibfield{author}{\bibinfo{person}{Shyama~Sastha Krishnamoorthy~Srinivasan}, \bibinfo{person}{Arhaan Bahadur}, \bibinfo{person}{Suruchi Singh}, \bibinfo{person}{Swati Kedia~gupta}, \bibinfo{person}{Vanya Jain}, \bibinfo{person}{Koushik~Sinha Deb}, \bibinfo{person}{Mohan Kumar}, {and} \bibinfo{person}{Pushpendra Singh}.} \bibinfo{year}{2025}\natexlab{}.
\newblock \showarticletitle{Demystifying Mental Health Reports Through an LLM-based Approach}. In \bibinfo{booktitle}{\emph{Proceedings of the Extended Abstracts of the CHI Conference on Human Factors in Computing Systems}} (Yokohama, Japan) \emph{(\bibinfo{series}{CHI EA '25})}. \bibinfo{publisher}{Association for Computing Machinery}, \bibinfo{address}{New York, NY, USA}, Article \bibinfo{articleno}{174}, \bibinfo{numpages}{7}~pages.
\newblock
\showISBNx{9798400713958}
\urldef\tempurl%
\url{https://doi.org/10.1145/3706599.3720208}
\showDOI{\tempurl}


\bibitem[Lee et~al\mbox{.}(2024)]%
        {10.1145/3637334}
\bibfield{author}{\bibinfo{person}{Hyunsoo Lee}, \bibinfo{person}{Yugyeong Jung}, \bibinfo{person}{Youwon Shin}, \bibinfo{person}{Hyesoo Park}, \bibinfo{person}{Woohyeok Choi}, {and} \bibinfo{person}{Uichin Lee}.} \bibinfo{year}{2024}\natexlab{}.
\newblock \showarticletitle{FamilyScope: Visualizing Affective Aspects of Family Social Interactions using Passive Sensor Data}.
\newblock \bibinfo{journal}{\emph{Proc. ACM Hum.-Comput. Interact.}} \bibinfo{volume}{8}, \bibinfo{number}{CSCW1}, Article \bibinfo{articleno}{57} (\bibinfo{date}{April} \bibinfo{year}{2024}), \bibinfo{numpages}{27}~pages.
\newblock
\urldef\tempurl%
\url{https://doi.org/10.1145/3637334}
\showDOI{\tempurl}


\bibitem[Li et~al\mbox{.}(2010)]%
        {10.1145/1753326.1753409}
\bibfield{author}{\bibinfo{person}{Ian Li}, \bibinfo{person}{Anind Dey}, {and} \bibinfo{person}{Jodi Forlizzi}.} \bibinfo{year}{2010}\natexlab{}.
\newblock \showarticletitle{A stage-based model of personal informatics systems}. In \bibinfo{booktitle}{\emph{Proceedings of the SIGCHI Conference on Human Factors in Computing Systems}} (Atlanta, Georgia, USA) \emph{(\bibinfo{series}{CHI '10})}. \bibinfo{publisher}{Association for Computing Machinery}, \bibinfo{address}{New York, NY, USA}, \bibinfo{pages}{557–566}.
\newblock
\showISBNx{9781605589299}
\urldef\tempurl%
\url{https://doi.org/10.1145/1753326.1753409}
\showDOI{\tempurl}


\bibitem[Liang et~al\mbox{.}(2023)]%
        {10.1145/3544549.3585604}
\bibfield{author}{\bibinfo{person}{Paul~Pu Liang}, \bibinfo{person}{Yiwei Lyu}, \bibinfo{person}{Gunjan Chhablani}, \bibinfo{person}{Nihal Jain}, \bibinfo{person}{Zihao Deng}, \bibinfo{person}{Xingbo Wang}, \bibinfo{person}{Louis-Philippe Morency}, {and} \bibinfo{person}{Ruslan Salakhutdinov}.} \bibinfo{year}{2023}\natexlab{}.
\newblock \showarticletitle{MultiViz: Towards User-Centric Visualizations and Interpretations of Multimodal Models}. In \bibinfo{booktitle}{\emph{Extended Abstracts of the 2023 CHI Conference on Human Factors in Computing Systems}} (Hamburg, Germany) \emph{(\bibinfo{series}{CHI EA '23})}. \bibinfo{publisher}{Association for Computing Machinery}, \bibinfo{address}{New York, NY, USA}, Article \bibinfo{articleno}{214}, \bibinfo{numpages}{21}~pages.
\newblock
\showISBNx{9781450394222}
\urldef\tempurl%
\url{https://doi.org/10.1145/3544549.3585604}
\showDOI{\tempurl}


\bibitem[Liao et~al\mbox{.}(2020)]%
        {10.1145/3313831.3376590}
\bibfield{author}{\bibinfo{person}{Q.~Vera Liao}, \bibinfo{person}{Daniel Gruen}, {and} \bibinfo{person}{Sarah Miller}.} \bibinfo{year}{2020}\natexlab{}.
\newblock \showarticletitle{Questioning the AI: Informing Design Practices for Explainable AI User Experiences}. In \bibinfo{booktitle}{\emph{Proceedings of the 2020 CHI Conference on Human Factors in Computing Systems}} (Honolulu, HI, USA) \emph{(\bibinfo{series}{CHI '20})}. \bibinfo{publisher}{Association for Computing Machinery}, \bibinfo{address}{New York, NY, USA}, \bibinfo{pages}{1–15}.
\newblock
\showISBNx{9781450367080}
\urldef\tempurl%
\url{https://doi.org/10.1145/3313831.3376590}
\showDOI{\tempurl}


\bibitem[Lim et~al\mbox{.}(2025)]%
        {10.1145/3706598.3714058}
\bibfield{author}{\bibinfo{person}{Brian~Y. Lim}, \bibinfo{person}{Joseph~P. Cahaly}, \bibinfo{person}{Chester Y.~F. Sng}, {and} \bibinfo{person}{Adam Chew}.} \bibinfo{year}{2025}\natexlab{}.
\newblock \showarticletitle{Diagrammatization and Abduction to Improve AI Interpretability With Domain-Aligned Explanations for Medical Diagnosis}. In \bibinfo{booktitle}{\emph{Proceedings of the 2025 CHI Conference on Human Factors in Computing Systems}} (Yokohama, Japan) \emph{(\bibinfo{series}{CHI '25})}. \bibinfo{publisher}{Association for Computing Machinery}, \bibinfo{address}{New York, NY, USA}, Article \bibinfo{articleno}{419}, \bibinfo{numpages}{25}~pages.
\newblock
\showISBNx{9798400713941}
\urldef\tempurl%
\url{https://doi.org/10.1145/3706598.3714058}
\showDOI{\tempurl}


\bibitem[Loerakker et~al\mbox{.}(2025)]%
        {10.1145/3706598.3713650}
\bibfield{author}{\bibinfo{person}{Meagan~B. Loerakker}, \bibinfo{person}{Tora Jarsve}, \bibinfo{person}{Jasmin Niess}, {and} \bibinfo{person}{Pawe\l{}~W. Wo\'{z}niak}.} \bibinfo{year}{2025}\natexlab{}.
\newblock \showarticletitle{The Framework of the Lived Experience of Metrics: Understanding the Purposes and Activities of Self-Tracking Metrics}. In \bibinfo{booktitle}{\emph{Proceedings of the 2025 CHI Conference on Human Factors in Computing Systems}} (Yokohama, Japan) \emph{(\bibinfo{series}{CHI '25})}. \bibinfo{publisher}{Association for Computing Machinery}, \bibinfo{address}{New York, NY, USA}, Article \bibinfo{articleno}{1189}, \bibinfo{numpages}{20}~pages.
\newblock
\showISBNx{9798400713941}
\urldef\tempurl%
\url{https://doi.org/10.1145/3706598.3713650}
\showDOI{\tempurl}


\bibitem[Luo et~al\mbox{.}(2025)]%
        {10.1145/3715336.3735746}
\bibfield{author}{\bibinfo{person}{Yuhan Luo}, \bibinfo{person}{Xinning Gui}, \bibinfo{person}{Xianghua(Sharon) Ding}, \bibinfo{person}{Xi Zheng}, \bibinfo{person}{Rie Helene~(Lindy) Hernandez}, \bibinfo{person}{Zhuoyang Li}, {and} \bibinfo{person}{Qiurong Song}.} \bibinfo{year}{2025}\natexlab{}.
\newblock \showarticletitle{Reflecting Upon The Unintended Consequences of Personal Informatics Systems: A Systematic Review of Empirical Studies}. In \bibinfo{booktitle}{\emph{Proceedings of the 2025 ACM Designing Interactive Systems Conference}} (Funchal, Portugal) \emph{(\bibinfo{series}{DIS '25})}. \bibinfo{publisher}{Association for Computing Machinery}, \bibinfo{address}{New York, NY, USA}, \bibinfo{pages}{2847–2866}.
\newblock
\showISBNx{9798400714856}
\urldef\tempurl%
\url{https://doi.org/10.1145/3715336.3735746}
\showDOI{\tempurl}


\bibitem[Mazurek et~al\mbox{.}(2010)]%
        {10.1145/1753326.1753421}
\bibfield{author}{\bibinfo{person}{Michelle~L. Mazurek}, \bibinfo{person}{J.~P. Arsenault}, \bibinfo{person}{Joanna Bresee}, \bibinfo{person}{Nitin Gupta}, \bibinfo{person}{Iulia Ion}, \bibinfo{person}{Christina Johns}, \bibinfo{person}{Daniel Lee}, \bibinfo{person}{Yuan Liang}, \bibinfo{person}{Jenny Olsen}, \bibinfo{person}{Brandon Salmon}, \bibinfo{person}{Richard Shay}, \bibinfo{person}{Kami Vaniea}, \bibinfo{person}{Lujo Bauer}, \bibinfo{person}{Lorrie~Faith Cranor}, \bibinfo{person}{Gregory~R. Ganger}, {and} \bibinfo{person}{Michael~K. Reiter}.} \bibinfo{year}{2010}\natexlab{}.
\newblock \showarticletitle{Access Control for Home Data Sharing: Attitudes, Needs and Practices}. In \bibinfo{booktitle}{\emph{Proceedings of the SIGCHI Conference on Human Factors in Computing Systems}} (Atlanta, Georgia, USA) \emph{(\bibinfo{series}{CHI '10})}. \bibinfo{publisher}{Association for Computing Machinery}, \bibinfo{address}{New York, NY, USA}, \bibinfo{pages}{645–654}.
\newblock
\showISBNx{9781605589299}
\urldef\tempurl%
\url{https://doi.org/10.1145/1753326.1753421}
\showDOI{\tempurl}


\bibitem[Molina-Markham et~al\mbox{.}(2014)]%
        {10.1145/2676431.2676432}
\bibfield{author}{\bibinfo{person}{Andres Molina-Markham}, \bibinfo{person}{Ronald Peterson}, \bibinfo{person}{Joseph Skinner}, \bibinfo{person}{Tianlong Yun}, \bibinfo{person}{Bhargav Golla}, \bibinfo{person}{Kevin Freeman}, \bibinfo{person}{Travis Peters}, \bibinfo{person}{Jacob Sorber}, \bibinfo{person}{Ryan Halter}, {and} \bibinfo{person}{David Kotz}.} \bibinfo{year}{2014}\natexlab{}.
\newblock \showarticletitle{Amulet: a secure architecture for mHealth applications for low-power wearable devices}. In \bibinfo{booktitle}{\emph{Proceedings of the 1st Workshop on Mobile Medical Applications}} (Memphis, Tennessee) \emph{(\bibinfo{series}{MMA '14})}. \bibinfo{publisher}{Association for Computing Machinery}, \bibinfo{address}{New York, NY, USA}, \bibinfo{pages}{16–21}.
\newblock
\showISBNx{9781450331906}
\urldef\tempurl%
\url{https://doi.org/10.1145/2676431.2676432}
\showDOI{\tempurl}


\bibitem[Moore et~al\mbox{.}(2025)]%
        {10.1145/3757599}
\bibfield{author}{\bibinfo{person}{Dylan~Edward Moore}, \bibinfo{person}{Songyun Tao}, \bibinfo{person}{Emma Ricci-De~Lucca}, \bibinfo{person}{Christina~Sapp Tadin}, \bibinfo{person}{Dio Tadin}, \bibinfo{person}{Brian Morgan}, {and} \bibinfo{person}{Elizabeth~L. Murnane}.} \bibinfo{year}{2025}\natexlab{}.
\newblock \showarticletitle{Family In The Loop: Enabling Family Involvement and Person-Centered Dementia Care at Long-Term Care Facilities with Collaborative AI Tools}.
\newblock \bibinfo{journal}{\emph{Proc. ACM Hum.-Comput. Interact.}} \bibinfo{volume}{9}, \bibinfo{number}{7}, Article \bibinfo{articleno}{CSCW418} (\bibinfo{date}{Oct.} \bibinfo{year}{2025}), \bibinfo{numpages}{45}~pages.
\newblock
\urldef\tempurl%
\url{https://doi.org/10.1145/3757599}
\showDOI{\tempurl}


\bibitem[Murnane et~al\mbox{.}(2018)]%
        {10.1145/3274396}
\bibfield{author}{\bibinfo{person}{Elizabeth~L. Murnane}, \bibinfo{person}{Tara~G. Walker}, \bibinfo{person}{Beck Tench}, \bibinfo{person}{Stephen Voida}, {and} \bibinfo{person}{Jaime Snyder}.} \bibinfo{year}{2018}\natexlab{}.
\newblock \showarticletitle{Personal Informatics in Interpersonal Contexts: Towards the Design of Technology that Supports the Social Ecologies of Long-Term Mental Health Management}.
\newblock \bibinfo{journal}{\emph{Proc. ACM Hum.-Comput. Interact.}} \bibinfo{volume}{2}, \bibinfo{number}{CSCW}, Article \bibinfo{articleno}{127} (\bibinfo{date}{Nov.} \bibinfo{year}{2018}), \bibinfo{numpages}{27}~pages.
\newblock
\urldef\tempurl%
\url{https://doi.org/10.1145/3274396}
\showDOI{\tempurl}


\bibitem[Nikkhah et~al\mbox{.}(2022)]%
        {10.1145/3555187}
\bibfield{author}{\bibinfo{person}{Sarah Nikkhah}, \bibinfo{person}{Swaroop John}, \bibinfo{person}{Krishna~Supradeep Yalamarti}, \bibinfo{person}{Emily~L. Mueller}, {and} \bibinfo{person}{Andrew~D. Miller}.} \bibinfo{year}{2022}\natexlab{}.
\newblock \showarticletitle{Family Care Coordination in the Children's Hospital: Phases and Cycles in the Pediatric Cancer Caregiving Journey}.
\newblock \bibinfo{journal}{\emph{Proc. ACM Hum.-Comput. Interact.}} \bibinfo{volume}{6}, \bibinfo{number}{CSCW2}, Article \bibinfo{articleno}{296} (\bibinfo{date}{Nov.} \bibinfo{year}{2022}), \bibinfo{numpages}{30}~pages.
\newblock
\urldef\tempurl%
\url{https://doi.org/10.1145/3555187}
\showDOI{\tempurl}


\bibitem[Nikkhah et~al\mbox{.}(2024)]%
        {10.1145/3686998}
\bibfield{author}{\bibinfo{person}{Sarah Nikkhah}, \bibinfo{person}{Akash~Uday Rode}, \bibinfo{person}{Neha~Keshav Kulkarni}, \bibinfo{person}{Priyanjali Mittal}, \bibinfo{person}{Emily~L. Mueller}, {and} \bibinfo{person}{Andrew~D. Miller}.} \bibinfo{year}{2024}\natexlab{}.
\newblock \showarticletitle{Family Resilience in Care Coordination Technologies: Designing for Families as Adaptive Systems}.
\newblock \bibinfo{journal}{\emph{Proc. ACM Hum.-Comput. Interact.}} \bibinfo{volume}{8}, \bibinfo{number}{CSCW2}, Article \bibinfo{articleno}{459} (\bibinfo{date}{Nov.} \bibinfo{year}{2024}), \bibinfo{numpages}{28}~pages.
\newblock
\urldef\tempurl%
\url{https://doi.org/10.1145/3686998}
\showDOI{\tempurl}


\bibitem[Nimmo et~al\mbox{.}(2024)]%
        {10.1145/3613904.3642352}
\bibfield{author}{\bibinfo{person}{Robert Nimmo}, \bibinfo{person}{Marios Constantinides}, \bibinfo{person}{Ke Zhou}, \bibinfo{person}{Daniele Quercia}, {and} \bibinfo{person}{Simone Stumpf}.} \bibinfo{year}{2024}\natexlab{}.
\newblock \showarticletitle{User Characteristics in Explainable AI: The Rabbit Hole of Personalization?}. In \bibinfo{booktitle}{\emph{Proceedings of the 2024 CHI Conference on Human Factors in Computing Systems}} (Honolulu, HI, USA) \emph{(\bibinfo{series}{CHI '24})}. \bibinfo{publisher}{Association for Computing Machinery}, \bibinfo{address}{New York, NY, USA}, Article \bibinfo{articleno}{317}, \bibinfo{numpages}{13}~pages.
\newblock
\showISBNx{9798400703300}
\urldef\tempurl%
\url{https://doi.org/10.1145/3613904.3642352}
\showDOI{\tempurl}


\bibitem[Nissenbaum(2004)]%
        {Nissenbaum_2004}
\bibfield{author}{\bibinfo{person}{Helen Nissenbaum}.} \bibinfo{year}{2004}\natexlab{}.
\newblock \showarticletitle{Privacy as Contextual Integrity}.
\newblock \bibinfo{journal}{\emph{Washington Law Review}} \bibinfo{volume}{79}, \bibinfo{number}{1} (\bibinfo{date}{Feb} \bibinfo{year}{2004}), \bibinfo{pages}{119–158}.
\newblock
\urldef\tempurl%
\url{https://digitalcommons.law.uw.edu/wlr/vol79/iss1/10/}
\showURL{%
\tempurl}


\bibitem[Oyebode et~al\mbox{.}(2021)]%
        {10.1145/3411764.3445619}
\bibfield{author}{\bibinfo{person}{Oladapo Oyebode}, \bibinfo{person}{Chinenye Ndulue}, \bibinfo{person}{Dinesh Mulchandani}, \bibinfo{person}{Ashfaq A.~Zamil~Adib}, \bibinfo{person}{Mona Alhasani}, {and} \bibinfo{person}{Rita Orji}.} \bibinfo{year}{2021}\natexlab{}.
\newblock \showarticletitle{Tailoring Persuasive and Behaviour Change Systems Based on Stages of Change and Motivation}. In \bibinfo{booktitle}{\emph{Proceedings of the 2021 CHI Conference on Human Factors in Computing Systems}} (Yokohama, Japan) \emph{(\bibinfo{series}{CHI '21})}. \bibinfo{publisher}{Association for Computing Machinery}, \bibinfo{address}{New York, NY, USA}, Article \bibinfo{articleno}{703}, \bibinfo{numpages}{19}~pages.
\newblock
\showISBNx{9781450380966}
\urldef\tempurl%
\url{https://doi.org/10.1145/3411764.3445619}
\showDOI{\tempurl}


\bibitem[Pham et~al\mbox{.}(2022)]%
        {10.1145/3495243.3560533}
\bibfield{author}{\bibinfo{person}{Nhat Pham}, \bibinfo{person}{Hong Jia}, \bibinfo{person}{Minh Tran}, \bibinfo{person}{Tuan Dinh}, \bibinfo{person}{Nam Bui}, \bibinfo{person}{Young Kwon}, \bibinfo{person}{Dong Ma}, \bibinfo{person}{Phuc Nguyen}, \bibinfo{person}{Cecilia Mascolo}, {and} \bibinfo{person}{Tam Vu}.} \bibinfo{year}{2022}\natexlab{}.
\newblock \showarticletitle{PROS: an efficient pattern-driven compressive sensing framework for low-power biopotential-based wearables with on-chip intelligence}. In \bibinfo{booktitle}{\emph{Proceedings of the 28th Annual International Conference on Mobile Computing And Networking}} (Sydney, NSW, Australia) \emph{(\bibinfo{series}{MobiCom '22})}. \bibinfo{publisher}{Association for Computing Machinery}, \bibinfo{address}{New York, NY, USA}, \bibinfo{pages}{661–675}.
\newblock
\showISBNx{9781450391818}
\urldef\tempurl%
\url{https://doi.org/10.1145/3495243.3560533}
\showDOI{\tempurl}


\bibitem[Pina et~al\mbox{.}(2017)]%
        {10.1145/2998181.2998362}
\bibfield{author}{\bibinfo{person}{Laura~R. Pina}, \bibinfo{person}{Sang-Wha Sien}, \bibinfo{person}{Teresa Ward}, \bibinfo{person}{Jason~C. Yip}, \bibinfo{person}{Sean~A. Munson}, \bibinfo{person}{James Fogarty}, {and} \bibinfo{person}{Julie~A. Kientz}.} \bibinfo{year}{2017}\natexlab{}.
\newblock \showarticletitle{From Personal Informatics to Family Informatics: Understanding Family Practices around Health Monitoring}. In \bibinfo{booktitle}{\emph{Proceedings of the 2017 ACM Conference on Computer Supported Cooperative Work and Social Computing}} (Portland, Oregon, USA) \emph{(\bibinfo{series}{CSCW '17})}. \bibinfo{publisher}{Association for Computing Machinery}, \bibinfo{address}{New York, NY, USA}, \bibinfo{pages}{2300–2315}.
\newblock
\showISBNx{9781450343350}
\urldef\tempurl%
\url{https://doi.org/10.1145/2998181.2998362}
\showDOI{\tempurl}


\bibitem[Pipek and Wulf(2009)]%
        {Pipek_Wulf_2009}
\bibfield{author}{\bibinfo{person}{Volkmar Pipek} {and} \bibinfo{person}{Volker Wulf}.} \bibinfo{year}{2009}\natexlab{}.
\newblock \showarticletitle{Infrastructuring: Toward an integrated perspective on the design and use of Information Technology}.
\newblock \bibinfo{journal}{\emph{Journal of the Association for Information Systems}} \bibinfo{volume}{10}, \bibinfo{number}{5} (\bibinfo{date}{May} \bibinfo{year}{2009}), \bibinfo{pages}{447–473}.
\newblock
\urldef\tempurl%
\url{https://doi.org/10.17705/1jais.00195}
\showDOI{\tempurl}


\bibitem[R\"{o}ddiger et~al\mbox{.}(2025)]%
        {10.1145/3712069}
\bibfield{author}{\bibinfo{person}{Tobias R\"{o}ddiger}, \bibinfo{person}{Michael K\"{u}ttner}, \bibinfo{person}{Philipp Lepold}, \bibinfo{person}{Tobias King}, \bibinfo{person}{Dennis Moschina}, \bibinfo{person}{Oliver Bagge}, \bibinfo{person}{Joseph~A. Paradiso}, \bibinfo{person}{Christopher Clarke}, {and} \bibinfo{person}{Michael Beigl}.} \bibinfo{year}{2025}\natexlab{}.
\newblock \showarticletitle{OpenEarable 2.0: Open-Source Earphone Platform for Physiological Ear Sensing}.
\newblock \bibinfo{journal}{\emph{Proc. ACM Interact. Mob. Wearable Ubiquitous Technol.}} \bibinfo{volume}{9}, \bibinfo{number}{1}, Article \bibinfo{articleno}{16} (\bibinfo{date}{March} \bibinfo{year}{2025}), \bibinfo{numpages}{33}~pages.
\newblock
\urldef\tempurl%
\url{https://doi.org/10.1145/3712069}
\showDOI{\tempurl}


\bibitem[Saksono et~al\mbox{.}(2020)]%
        {10.5555/AAI10223295}
\bibfield{author}{\bibinfo{person}{Herman Saksono}, \bibinfo{person}{Stephen Intille}, \bibinfo{person}{Magy~Seif El-Nasr}, {and} \bibinfo{person}{Sean Munson}.} \bibinfo{year}{2020}\natexlab{}.
\newblock \emph{\bibinfo{title}{A Social Cognition Framework for Interpersonal Informatics in Families}}.
\newblock \bibinfo{thesistype}{Ph.\,D. Dissertation}. \bibinfo{school}{Northeastern University}, \bibinfo{address}{USA}.
\newblock Advisor(s) G, Parker, Andrea.
\newblock
\showISBNx{9798664745634}
\newblock
\shownote{AAI10223295}.


\bibitem[Saleem and Jan(2021)]%
        {saleemmodifiedkuppuswamysocioeconomic2021}
\bibfield{author}{\bibinfo{person}{Sheikh~Mohd Saleem} {and} \bibinfo{person}{Shah~Sumaya Jan}.} \bibinfo{year}{2021}\natexlab{}.
\newblock \showarticletitle{Modified {{Kuppuswamy}} Socioeconomic Scale Updated for the Year 2021}.
\newblock \bibinfo{journal}{\emph{Indian Journal of Forensic and Community Medicine}} \bibinfo{volume}{8}, \bibinfo{number}{1} (\bibinfo{date}{March} \bibinfo{year}{2021}), \bibinfo{pages}{1--3}.
\newblock
\showISSN{2394-6776}
\urldef\tempurl%
\url{https://doi.org/10.18231/j.ijfcm.2021.001}
\showDOI{\tempurl}


\bibitem[Shin et~al\mbox{.}(2022)]%
        {10.1145/3491102.3502041}
\bibfield{author}{\bibinfo{person}{Jaemin Shin}, \bibinfo{person}{Seungjoo Lee}, \bibinfo{person}{Taesik Gong}, \bibinfo{person}{Hyungjun Yoon}, \bibinfo{person}{Hyunchul Roh}, \bibinfo{person}{Andrea Bianchi}, {and} \bibinfo{person}{Sung-Ju Lee}.} \bibinfo{year}{2022}\natexlab{}.
\newblock \showarticletitle{MyDJ: Sensing Food Intakes with an Attachable on Your Eyeglass Frame}. In \bibinfo{booktitle}{\emph{Proceedings of the 2022 CHI Conference on Human Factors in Computing Systems}} (New Orleans, LA, USA) \emph{(\bibinfo{series}{CHI '22})}. \bibinfo{publisher}{Association for Computing Machinery}, \bibinfo{address}{New York, NY, USA}, Article \bibinfo{articleno}{341}, \bibinfo{numpages}{17}~pages.
\newblock
\showISBNx{9781450391573}
\urldef\tempurl%
\url{https://doi.org/10.1145/3491102.3502041}
\showDOI{\tempurl}


\bibitem[Shin et~al\mbox{.}(2019)]%
        {10.1145/3307334.3328565}
\bibfield{author}{\bibinfo{person}{Jaemin Shin}, \bibinfo{person}{Seungjoo Lee}, {and} \bibinfo{person}{Sung-Ju Lee}.} \bibinfo{year}{2019}\natexlab{}.
\newblock \showarticletitle{Accurate Eating Detection on a Daily Wearable Necklace (demo)}. In \bibinfo{booktitle}{\emph{Proceedings of the 17th Annual International Conference on Mobile Systems, Applications, and Services}} (Seoul, Republic of Korea) \emph{(\bibinfo{series}{MobiSys '19})}. \bibinfo{publisher}{Association for Computing Machinery}, \bibinfo{address}{New York, NY, USA}, \bibinfo{pages}{649–650}.
\newblock
\showISBNx{9781450366618}
\urldef\tempurl%
\url{https://doi.org/10.1145/3307334.3328565}
\showDOI{\tempurl}


\bibitem[Shove et~al\mbox{.}(2014)]%
        {shovesocial}
\bibfield{author}{\bibinfo{person}{Elizabeth Shove}, \bibinfo{person}{Mika Pantzar}, \bibinfo{person}{Matt Watson}, {et~al\mbox{.}}} \bibinfo{year}{2014}\natexlab{}.
\newblock \showarticletitle{Making and Breaking Links}.
\newblock In \bibinfo{booktitle}{\emph{The Dynamics of Social Practice: Everyday Life and How it Changes}}. \bibinfo{publisher}{SAGE Publications Ltd}, \bibinfo{address}{1 Oliver's Yard, 55 City Road, London EC1Y 1SP United Kingdom}, \bibinfo{pages}{21--42}.
\newblock


\bibitem[Soubutts et~al\mbox{.}(2023)]%
        {10.1145/3544548.3581546}
\bibfield{author}{\bibinfo{person}{Ewan Soubutts}, \bibinfo{person}{Elaine Czech}, \bibinfo{person}{Amid Ayobi}, \bibinfo{person}{Rachel Eardley}, \bibinfo{person}{Kirsten Cater}, {and} \bibinfo{person}{Aisling~Ann O'Kane}.} \bibinfo{year}{2023}\natexlab{}.
\newblock \showarticletitle{The Shifting Sands of Labour: Changes in Shared Care Work with a Smart Home Health System}. In \bibinfo{booktitle}{\emph{Proceedings of the 2023 CHI Conference on Human Factors in Computing Systems}} (Hamburg, Germany) \emph{(\bibinfo{series}{CHI '23})}. \bibinfo{publisher}{Association for Computing Machinery}, \bibinfo{address}{New York, NY, USA}, Article \bibinfo{articleno}{490}, \bibinfo{numpages}{16}~pages.
\newblock
\showISBNx{9781450394215}
\urldef\tempurl%
\url{https://doi.org/10.1145/3544548.3581546}
\showDOI{\tempurl}


\bibitem[Srinivasan et~al\mbox{.}(2026)]%
        {srinivasan2026unpackingpersonalhealthinformatics}
\bibfield{author}{\bibinfo{person}{Shyama Sastha~Krishnamoorthy Srinivasan}, \bibinfo{person}{Mohan Kumar}, {and} \bibinfo{person}{Pushpendra Singh}.} \bibinfo{year}{2026}\natexlab{}.
\newblock \bibinfo{title}{Unpacking "Personal" Health Informatics for Proactive Collective Care}.
\newblock
\newblock
\showeprint[arxiv]{2509.01231}~[cs.HC]
\urldef\tempurl%
\url{https://arxiv.org/abs/2509.01231}
\showURL{%
\tempurl}


\bibitem[Star and Ruhleder(1996)]%
        {Star_Ruhleder_1996}
\bibfield{author}{\bibinfo{person}{Susan~Leigh Star} {and} \bibinfo{person}{Karen Ruhleder}.} \bibinfo{year}{1996}\natexlab{}.
\newblock \showarticletitle{Steps toward an ecology of infrastructure: Design and access for large information spaces}.
\newblock \bibinfo{journal}{\emph{Information Systems Research}} \bibinfo{volume}{7}, \bibinfo{number}{1} (\bibinfo{date}{Mar} \bibinfo{year}{1996}), \bibinfo{pages}{111–134}.
\newblock
\urldef\tempurl%
\url{https://doi.org/10.1287/isre.7.1.111}
\showDOI{\tempurl}


\bibitem[Toombs et~al\mbox{.}(2018)]%
        {10.1145/3197391.3197394}
\bibfield{author}{\bibinfo{person}{Austin~L. Toombs}, \bibinfo{person}{Andy Dow}, \bibinfo{person}{John Vines}, \bibinfo{person}{Colin~M. Gray}, \bibinfo{person}{Barbara Dennis}, \bibinfo{person}{Rachel Clarke}, {and} \bibinfo{person}{Ann Light}.} \bibinfo{year}{2018}\natexlab{}.
\newblock \showarticletitle{Designing for Everyday Care in Communities}. In \bibinfo{booktitle}{\emph{Proceedings of the 2018 ACM Conference Companion Publication on Designing Interactive Systems}} (Hong Kong, China) \emph{(\bibinfo{series}{DIS '18 Companion})}. \bibinfo{publisher}{Association for Computing Machinery}, \bibinfo{address}{New York, NY, USA}, \bibinfo{pages}{391–394}.
\newblock
\showISBNx{9781450356312}
\urldef\tempurl%
\url{https://doi.org/10.1145/3197391.3197394}
\showDOI{\tempurl}


\bibitem[Truong et~al\mbox{.}(2020)]%
        {10.1145/3386901.3389022}
\bibfield{author}{\bibinfo{person}{Hoang Truong}, \bibinfo{person}{Nam Bui}, \bibinfo{person}{Zohreh Raghebi}, \bibinfo{person}{Marta Ceko}, \bibinfo{person}{Nhat Pham}, \bibinfo{person}{Phuc Nguyen}, \bibinfo{person}{Anh Nguyen}, \bibinfo{person}{Taeho Kim}, \bibinfo{person}{Katrina Siegfried}, \bibinfo{person}{Evan Stene}, \bibinfo{person}{Taylor Tvrdy}, \bibinfo{person}{Logan Weinman}, \bibinfo{person}{Thomas Payne}, \bibinfo{person}{Devin Burke}, \bibinfo{person}{Thang Dinh}, \bibinfo{person}{Sidney D'Mello}, \bibinfo{person}{Farnoush Banaei-Kashani}, \bibinfo{person}{Tor Wager}, \bibinfo{person}{Pavel Goldstein}, {and} \bibinfo{person}{Tam Vu}.} \bibinfo{year}{2020}\natexlab{}.
\newblock \showarticletitle{Painometry: wearable and objective quantification system for acute postoperative pain}. In \bibinfo{booktitle}{\emph{Proceedings of the 18th International Conference on Mobile Systems, Applications, and Services}} (Toronto, Ontario, Canada) \emph{(\bibinfo{series}{MobiSys '20})}. \bibinfo{publisher}{Association for Computing Machinery}, \bibinfo{address}{New York, NY, USA}, \bibinfo{pages}{419–433}.
\newblock
\showISBNx{9781450379540}
\urldef\tempurl%
\url{https://doi.org/10.1145/3386901.3389022}
\showDOI{\tempurl}


\bibitem[Wang et~al\mbox{.}(2023)]%
        {10.1145/3563657.3596096}
\bibfield{author}{\bibinfo{person}{Weiyun Wang}, \bibinfo{person}{Xianghua(Sharon) Ding}, {and} \bibinfo{person}{Ilyena Hirskyj-Douglas}.} \bibinfo{year}{2023}\natexlab{}.
\newblock \showarticletitle{Everyday Space as an Interface for Health Data Engagement: Designing Tangible Displays of Stress Data}. In \bibinfo{booktitle}{\emph{Proceedings of the 2023 ACM Designing Interactive Systems Conference}} (Pittsburgh, PA, USA) \emph{(\bibinfo{series}{DIS '23})}. \bibinfo{publisher}{Association for Computing Machinery}, \bibinfo{address}{New York, NY, USA}, \bibinfo{pages}{1648–1659}.
\newblock
\showISBNx{9781450398930}
\urldef\tempurl%
\url{https://doi.org/10.1145/3563657.3596096}
\showDOI{\tempurl}


\bibitem[Wong et~al\mbox{.}(2021)]%
        {10.1145/3461778.3462046}
\bibfield{author}{\bibinfo{person}{Joshua Wong}, \bibinfo{person}{Pin~Sym Foong}, {and} \bibinfo{person}{Alex Mitchell}.} \bibinfo{year}{2021}\natexlab{}.
\newblock \showarticletitle{Contemplative Interactions: Exploring the Use of Defamiliarization in a Serious Game to Promote Reflective Thinking about Personal Health}. In \bibinfo{booktitle}{\emph{Proceedings of the 2021 ACM Designing Interactive Systems Conference}} (Virtual Event, USA) \emph{(\bibinfo{series}{DIS '21})}. \bibinfo{publisher}{Association for Computing Machinery}, \bibinfo{address}{New York, NY, USA}, \bibinfo{pages}{984–998}.
\newblock
\showISBNx{9781450384766}
\urldef\tempurl%
\url{https://doi.org/10.1145/3461778.3462046}
\showDOI{\tempurl}


\bibitem[Zhang et~al\mbox{.}(2020a)]%
        {10.1145/3386901.3389028}
\bibfield{author}{\bibinfo{person}{Hanbin Zhang}, \bibinfo{person}{Gabriel Guo}, \bibinfo{person}{Emery Comstock}, \bibinfo{person}{Baicheng Chen}, \bibinfo{person}{Xingyu Chen}, \bibinfo{person}{Chen Song}, \bibinfo{person}{Jerry Ajay}, \bibinfo{person}{Jeanne Langan}, \bibinfo{person}{Sutanuka Bhattacharjya}, \bibinfo{person}{Lora~A Cavuoto}, {and} \bibinfo{person}{Wenyao Xu}.} \bibinfo{year}{2020}\natexlab{a}.
\newblock \showarticletitle{RehabPhone: a software-defined tool using 3D printing and smartphones for personalized home-based rehabilitation}. In \bibinfo{booktitle}{\emph{Proceedings of the 18th International Conference on Mobile Systems, Applications, and Services}} (Toronto, Ontario, Canada) \emph{(\bibinfo{series}{MobiSys '20})}. \bibinfo{publisher}{Association for Computing Machinery}, \bibinfo{address}{New York, NY, USA}, \bibinfo{pages}{434–447}.
\newblock
\showISBNx{9781450379540}
\urldef\tempurl%
\url{https://doi.org/10.1145/3386901.3389028}
\showDOI{\tempurl}


\bibitem[Zhang et~al\mbox{.}(2020b)]%
        {10.1145/3397313}
\bibfield{author}{\bibinfo{person}{Shibo Zhang}, \bibinfo{person}{Yuqi Zhao}, \bibinfo{person}{Dzung~Tri Nguyen}, \bibinfo{person}{Runsheng Xu}, \bibinfo{person}{Sougata Sen}, \bibinfo{person}{Josiah Hester}, {and} \bibinfo{person}{Nabil Alshurafa}.} \bibinfo{year}{2020}\natexlab{b}.
\newblock \showarticletitle{NeckSense: A Multi-Sensor Necklace for Detecting Eating Activities in Free-Living Conditions}.
\newblock \bibinfo{journal}{\emph{Proc. ACM Interact. Mob. Wearable Ubiquitous Technol.}} \bibinfo{volume}{4}, \bibinfo{number}{2}, Article \bibinfo{articleno}{72} (\bibinfo{date}{June} \bibinfo{year}{2020}), \bibinfo{numpages}{26}~pages.
\newblock
\urldef\tempurl%
\url{https://doi.org/10.1145/3397313}
\showDOI{\tempurl}


\bibitem[Zhu et~al\mbox{.}(2025a)]%
        {10.1145/3706598.3714072}
\bibfield{author}{\bibinfo{person}{Jun Zhu}, \bibinfo{person}{Sruzan Lolla}, \bibinfo{person}{Meeshu Agnihotri}, \bibinfo{person}{Sahar Asgari~Tappeh}, \bibinfo{person}{Lala Guluzade}, \bibinfo{person}{Elena Agapie}, {and} \bibinfo{person}{Corina Sas}.} \bibinfo{year}{2025}\natexlab{a}.
\newblock \showarticletitle{A Systematic Review and Meta-Analysis of Research on Goals for Behavior Change}. In \bibinfo{booktitle}{\emph{Proceedings of the 2025 CHI Conference on Human Factors in Computing Systems}} (Yokohama, Japan) \emph{(\bibinfo{series}{CHI '25})}. \bibinfo{publisher}{Association for Computing Machinery}, \bibinfo{address}{New York, NY, USA}, Article \bibinfo{articleno}{375}, \bibinfo{numpages}{25}~pages.
\newblock
\showISBNx{9798400713941}
\urldef\tempurl%
\url{https://doi.org/10.1145/3706598.3714072}
\showDOI{\tempurl}


\bibitem[Zhu et~al\mbox{.}(2025b)]%
        {10.1145/3715668.3736372}
\bibfield{author}{\bibinfo{person}{Pengyu Zhu}, \bibinfo{person}{Yiming Yao}, \bibinfo{person}{Haiyue Lu}, \bibinfo{person}{Tse~Pei Ng}, \bibinfo{person}{Celeste Seah}, \bibinfo{person}{Janghee Cho}, \bibinfo{person}{Yiying Wu}, {and} \bibinfo{person}{Jung-Joo Lee}.} \bibinfo{year}{2025}\natexlab{b}.
\newblock \bibinfo{booktitle}{\emph{Unveiling the Invisible Work of Migrant Care Workers in Aged Care Homes: Implications for Future Care Technologies}}.
\newblock \bibinfo{publisher}{Association for Computing Machinery}, \bibinfo{address}{New York, NY, USA}, \bibinfo{pages}{329–334}.
\newblock
\showISBNx{9798400714863}
\urldef\tempurl%
\url{https://doi.org/10.1145/3715668.3736372}
\showURL{%
\tempurl}


\end{thebibliography}
